\RequirePackage{fix-cm}
\documentclass[smallextended]{svjour3}       
\smartqed  
\usepackage{graphicx}
\usepackage{verbatim}
\usepackage{amsfonts}
\usepackage{hyperref}
\usepackage{amssymb}
\usepackage{color}
\usepackage{amsmath}

\def\vzeta{\vec{\boldsymbol\zeta}}

\newtheorem{prop}{Proposition}

\begin{document}

\title{A set of invariant quality factors measuring the deviation from the Kerr metric
}

\titlerunning{A set of invariant quality factors measuring the deviation from the Kerr metric}        

\author{Alfonso Garc\'{i}a-Parrado G\'{o}mez-Lobo        \and
        Jos\'e M. M. Senovilla 
}

\authorrunning{A. Garc\'{\i}a-Parrado and J. M. M. Senovilla} 

\institute{Alfonso Garc\'{i}a-Parrado G\'{o}mez-Lobo \at
              Centro de Matem\'atica, Universidade do Minho\\ 
              4710-057 Braga, Portugal\\
              \email{alfonso@math.uminho.pt}           
           \and
           Jos\'e M. M. Senovilla \at
           F\'{\i}sica Te\'orica, Universidad del Pa\'{\i}s Vasco \\ 
           Apartado 644, E-48080 Bilbao, Spain\\
	    \email{josemm.senovilla@ehu.es}
}

\date{Received: date / Accepted: date}

\maketitle

\begin{abstract}
A number of scalar invariant characterizations of the Kerr solution are presented. These characterizations come in the form of 
{\em quality factors} defined in stationary space-times. A quality factor is a scalar quantity 
varying in the interval $[0,1]$ with the value 1 being attained if and only if the space-time is locally isometric 
to the Kerr solution. No knowledge of the Kerr solution is required to compute these quality factors.
A number of different possibilities arise depending on whether the space-time is Ricci-flat and asymptotically flat, 
just Ricci-flat, or Ricci non-flat. In each situation a number of 
quality factors are constructed and analysed.  
The relevance of these quality factors is clear in any situation where one seeks a 
rigorous formulation of the statement that a space-time is ``close'' to the Kerr solution, such as: 
its non-linear stability problem, the asymptotic settlement of a radiating isolated system undergoing gravitational collapse, or in the formulation
of some uniqueness results.
\end{abstract}

\section{Introduction}
The Kerr metric \cite{KERR-METRIC} is a fundamental exact solution in General Relativity providing the unique set of models describing the region outside a rotating black hole, so that its physical relevance is 
unquestionable. The uniqueness properties of the Kerr solution stem from powerful mathematical results stating, roughly speaking, that it is 
the only analytic, asymptotically flat vacuum solution of the Einstein field equations (without cosmological constant) representing a black hole with a {\em regular} event horizon 
(a precise, up-to-date formulation of this uniqueness result and its proof can be found
in \cite{CHRUSCIEL-COSTA,LRR-2012-7}). The generalization of this result for non-analytic solutions of the field equations
has been an open question for many years and recent progress towards its solution has been reported in 
\cite{IONESCU-KLAINERMAN,IONESCU-KL-ALEXAKIS,IONESCU-KLAINERMAN-K}   

A completely open problem regarding the Kerr solution is its {\em stability} under {\em nonlinear} perturbations. The non-linear stability of 
the Kerr solution ---to be more precise, the non-linear stability of its {\em domain of outer communication}--- 
can be formulated as the global existence of a solution of the vacuum Einstein equations corresponding to initial data whose maximal Cauchy development is {\em close} to a subset of the Kerr's domain of outer communication. Results 
in this direction already exist for other well known exact solutions to the Einstein field equations 
\cite{FRI86A,FRI86B,FRI91,CHRKLA93,KLAINNICO03,LINROD05} and therefore, given the relevance of the Kerr solution, 
it is important attempting to obtain
similar results for it.  In this regard, it is crucial to establish what is the precise meaning of the statement that a solution of the 
field equations be ``close to the Kerr solution''. 

Another point of relevance of the Kerr solution arises from its status as the stationary equilibrium limit reached in the evolution
of an isolated system undergoing gravitational collapse. Due to its uniqueness properties just mentioned, it is widely believed that after 
the gravitational wave content has been radiated away such systems will eventually settle down as the exterior region of a Kerr solution. 
Any rigorous proof of this statement requires to introduce the notion of ``approaching the 
(exterior) of the Kerr solution'' which in turn relies on a further notion of ``being close to Kerr''. With the aid of
this notion one could devise numerical 
simulations of a radiating isolated system and test whether in the asymptotic regime it is close to the Kerr solution. Note that the hypothesis just
mentioned that the Kerr metric is the asymptotic limit of a gravitational collapse process assumes implicitly the global existence of a solution 
``close'' to Kerr and therefore the non-linear stability of the Kerr solution discussed previously. Therefore if the non-linear stability
of the Kerr solution fails then the Kerr solution could not be thought of as the asymptotic limit in any realistic physical process.

The purpose of the present work is to put forward a number of {\em local and invariant} criteria which enable us to 
answer the question of when a given stationary space-time is ``close'' to the Kerr solution. By local and invariant we mean that the 
criteria are formulated in terms of scalar quantities, so that they have a perfectly defined value at each point of the space-time, and are independent of the coordinate systems used. This makes them suitable 
for their application to the problems discussed in the previous paragraphs. Our approach consists in introducing 
{\em quality factors} to measure the deviation of a given solution from the Kerr space-time. A quality factor is a 
dimensionless scalar quantity varying in the interval $[0,1]$ and such that the unity value is attained on a neighborhood of a point if and only if the space-time  is isometric to the Kerr solution there. Ideally one seeks quality factors which 
are constructed from invariantly 
defined quantities of the geometry. The existence of such type of quality factors permits to quantify invariantly, 
by a normalized non-negative real number, ``how close'' to the Kerr solution the space-time under consideration is. 

It is interesting to compare our approach with the work presented in 
\cite{KERR-INVARIANT-PRL,KERR-INVARIANT-TJ,KERR-INVARIANT-PRS,NON-KERRNESS-COMPACT}. In these references the authors present 
a positive scalar quantity (that they call {\em non-Kerrness})
constructed from a vacuum initial data set with certain properties. The invariant vanishes if and only if the data correspond to Kerr initial
data. If the Kerr non-linear stability were true, then one could expect that the maximal development of an initial 
data set close to Kerr initial data would also be close to the Kerr solution itself. In this sense the quality factors presented in this work
and the non-Kerrness studied in the above references could be related.

Our quality factors are obtained from a number of coordinate-free 
cha\-rac\-terisations of the Kerr solution.
A coordinate-free characterisation of the Kerr solution is a set of geometric conditions on a space-time 
which are fulfilled if and only if the space-time is locally isometric to the Kerr
solution. A number of them have been presented in the literature, and we rely specifically on those
which are formulated as local conditions, computable at any given point. An example of 
these can be found in \cite{FERSAEZKERR} where the Kerr metric is characterised in terms of a set of conditions on the concomitants of the curvature tensor at a point, so they are fully invariant. 
Another local invariant characterisation of the Kerr solution was put forward in \cite{MARS-KERR}. In this case one 
assumes the existence of a Killing vector in the space-time (an invariant property) and that the space-time is 
a vacuum solution. The characterization is then written as a set of local conditions involving the Killing vector 
and the curvature tensor. This result is the 
starting point of our work. The plan is to cast this invariant characterisation as a single scalar condition 
having the properties of a quality factor. Indeed, it turns out that this can be done in more than one way and
we present a number of quality factors which are relevant in different situations. 

An important property of our approach is that the quality factors that we introduce depend 
exclusively on invariant geometric properties of the space-time under study. 
One does not need to know anything about the Kerr solution, nor one has to put the space-time 
in relation with the Kerr solution (by diffeomorphisms, for instance). 
Actually, the quality factors can be explicitly computed for any given stationary space-time.

The question of whether or not our quality factors are good enough to measure the deviation 
of solutions from the Kerr metric will have to be evaluated, and a calibration 
of the meaning of any quality factor having values close to one (say 0.95 or more) should be carefully 
and exhaustively performed. In this sense, we support our analysis with the study of some practical 
examples where the quality factors have been explicitly computed. 
The results are in agreement with intuitive expectations of when a space-time can be considered close to the Kerr solution.

The plan of this work is as follows: in section \ref{sec:preliminaries} we recall some generalities of vacuum
space-times possessing a Killing vector. In sections \ref{sec:mars-simon-tensor} and \ref{sec:superenergy-mars-simon}
we construct a quality factor valid for any stationary, Ricci-flat and asymptotically flat space-time. This quality factor is based
on the so-called {\em Mars-Simon} tensor. The more general case of stationary
space-times is addressed in section \ref{sec:qfactors} where we construct a number of quality factors based on
the spacetime {\em Simon tensor}. We discuss quality factors valid for vacuum solutions and a quality factor which applies to 
non-vacuum solutions (although in this case the quality factor being unity is just a necessary condition 
for the space-time to be locally isometric to the Kerr solution). In section \ref{sec:examples}, we
carried out numerical computations of the quality factors in practical cases  
with the aid of the system {\em xAct} \cite{XACT,XPERM}. 
Finally we discuss some open issues of our analysis in section \ref{sec:conclusions}.

\section{Geometric preliminaries}
\label{sec:preliminaries}

Let $\mathcal M$ be a 4-dimensional smooth manifold 
endowed with a smooth Lorentzian metric $g_{ab}$ (signature convention $(-,+,+,+)$). 
The pair $({\mathcal M},g_{ab})$ will be referred to as a {\em Lorentzian manifold} or 
{\em spacetime}. Small Latin letters $a,b,c,\dots$ will be used to denote abstract indices for tensors fields 
on any tensor bundle built from $T(\mathcal M)$ and its dual $T^*(\mathcal M)$. Square (rounded) brackets 
enclosing a group of indices will denote the antisymmetrization (symmetrization) operation. The inverse of
 $g_{ab}$ is $g^{ab}$ and these quantities can be used to raise and lower tensor indices in the standard way. 
The Levi-Civita covariant derivative compatible with $g_{ab}$ is the operator $\nabla_a$, while 
$R_{abc}^{\phantom{abc}d}$ and $R_{ab}$ denote respectively the Riemann and Ricci tensors arising from 
$\nabla_a$. Our conventions in the definitions of these quantities are set by the relations,
$$
\nabla_a\nabla_b Z^c-\nabla_b\nabla_a Z^c=R_{bad}^{\phantom{bad}c}Z^d\;,\quad
R_{ab}\equiv R_{acb}^{\phantom{acb}c},
$$ 
where the first relation is just the Ricci identity valid for any vector field $Z^c$. 
Sometimes we will denote tensorial quantities with index-free notation in which case 
boldface capital letters will be used (the rank of the tensorial quantity should then be clear from the context). 
The complex conjugate of a complex quantity will be denoted with an overbar. 

A vector field $\vec{\boldsymbol\zeta}$ on ${\mathcal M}$ is called a {\em Killing vector} if it fulfills the differential condition 
\begin{equation}
\pounds_{\vec{\boldsymbol\zeta}} g_{ab}=0. 
\label{eq:killing-cond}
\end{equation}
where $\pounds_{\vec{\boldsymbol\zeta}}$ denotes the Lie derivative with respect to 
$\vec{\boldsymbol\zeta}$. From now on we will assume that the spacetime on which we work possesses a Killing vector. 
Killing vectors have been extensively studied and their properties can be found in many places of the literature but in order 
to make our presentation as self-contained as possible we will review here those properties needed in our work. 
First of all we assume that the Killing vector $\vec{\boldsymbol\zeta}$ is a smooth vector field in $\mathcal M$ 
(indeed, the smoothness of the Killing vector follows if one just assumes that it is a $C^2$ vector field, see below). 
The condition (\ref{eq:killing-cond}) implies that $F_{ab}\equiv \nabla_{a}\zeta_{b}$ is antisymmetric and closed when regarded as a 2-form
(Killing 2-form). Using the Killing condition (\ref{eq:killing-cond}) and the Ricci identity one may 
compute the covariant derivative of the Killing 2-form, yielding
\begin{equation}
\nabla_b F_{ac}=-R_{acbd}\zeta^d. 
\label{eq:cdfz}
\end{equation}

The norm of the Killing vector is $\lambda\equiv\zeta_a\zeta^a$ and no restriction on it 
is imposed at this stage. 
The properties discussed so far hold regardless of the space-time dimension and its matter content. 
In the particular case of Ricci-flat 4-dimensional space-time we have additional properties to be detailed next. 
First of all the Riemann tensor coincides 
with the Weyl tensor $C_{abcd}$ which in four dimensions has the algebraic properties $C^*_{abcd}=\ ^*C_{abcd}$ where 
the Weyl tensor right and left duals are defined respectively by
$$
C^*_{abcd}\equiv \frac{1}{2}\eta_{cdpq}C_{ab}^{\phantom{ab}pq}\;,\quad
\ ^*C_{abcd}\equiv \frac{1}{2}\eta_{abpq}C_{ab}^{\phantom{ab}pq}\;,
$$
and $\eta_{abcd}$ is the canonical volume 4-form. Therefore, we will only use the Weyl tensor and its right dual from now on. 
Also, in four dimensions one can define the dual of the Killing 2-form $F_{ab}$ by $F^*_{ab}\equiv F^{pq}\eta_{abpq}/2$. 
From $F^*_{ab}$ we introduce the
twist 1-form $\omega_a\equiv F^*_{ab}\zeta^b$ which, for a Ricci-flat spacetime, can be shown to be a closed 1-form, 
$\nabla_{[a}\omega_{b]}=0$. The {\em local} potential arising from $\omega_a$ 
is called the {\em twist potential} $\omega$. 
It is possible to achieve a great simplification in many computations by introducing {\em complex quantities}. 
In this way we define the {\em self-dual} Weyl tensor, and the self-dual Killing 2-form by the expressions
\begin{equation}
{\mathcal C}_{abcd}\equiv C_{abcd}+\mbox{i}\;C^*_{abcd}\;,\quad
{\mathcal F}_{ab}\equiv F_{ab}+\mbox{i}F^*_{ab}. 
\label{eq:self-dual-weyl-f}
\end{equation}
These quantities fulfill the well-known algebraic properties
\begin{equation}
\mathcal{C}^*_{abpq}\equiv\frac{1}{2} \eta_{pqcd} \mathcal{C}_{ab}{}^{cd} = -\mbox{i}\;\mathcal{C}_{abpq}\;,\quad
\mathcal{F}^*_{ab}\equiv\frac{1}{2} \eta_{abcd}\mathcal{F}^{cd}=-\mbox{i}\;\mathcal{F}_{ab}
\end{equation}
and the identities
\begin{equation}
{\mathcal F}_{ac}{\mathcal F}_b{}^c = \frac{1}{4} ({\mathcal F}\cdot{\mathcal F})g_{ab}\;,\quad
{\mathcal F}_{ac}\overline{\mathcal F}_b{}^c={\mathcal F}_{bc}\overline{\mathcal F}_a{}^c\;,\quad
{\mathcal F}\cdot{\mathcal F}\equiv{\mathcal F}_{ab}{\mathcal F}^{ab}\, .
\label{eq:F^2}
\end{equation}
With the aid of these new quantities, one can write the differential conditions fulfilled by $F_{ab}$ and $F^*_{ab}$ 
in a compact form
\begin{equation}
\nabla_{c}{\mathcal F}_{ab}=-{\mathcal C}_{abcd}\zeta^d\;,\quad
\nabla_{[c}{\mathcal F}_{ab]}=0. 
\label{eq:maxwell}
\end{equation}
Also, if we introduce the Ernst 1-form 
$$\sigma_a\equiv{2\mathcal F}_{ab}\zeta^b, \hspace{1cm} (\sigma_a\zeta^a =0)$$
one has that in a Ricci-flat space-time
\begin{equation}
\nabla_{[a}\sigma_{b]}=0\;, 
\end{equation}
which means that one can find a local potential  $\sigma$ for $\sigma_a$, called the Ernst potential, which can be defined in terms of 
the Killing norm and twist as 
\begin{equation}
\sigma\equiv\lambda+2\;\mbox{i}\;\omega\Rightarrow \nabla_a\sigma=\sigma_a. 
\label{eq:sigma}
\end{equation}
More generally, the Ernst 1-form is known to be closed in any space-time that satisfies $\zeta_{[c}R_{a]b}\zeta^b=0$.
Observe that $\sigma$ is defined up to the addition of an additive complex constant, so that we have the gauge freedom
\begin{equation}
\sigma \longrightarrow \sigma' =\sigma +\alpha\;,\quad \alpha \in \mathbb{C}\, .
\label{eq:free}
\end{equation}
This freedom will be relevant later when we deal with spacetimes locally, without reference to their possible asymptotic properties.
In this work by ``Ernst potential'' we mean any specific choice of type (\ref{eq:sigma}) for the potential of $\sigma_a$, and this will always be denoted by $\sigma$.
The quantities $\mathcal{F}_{ab}$ and $\sigma_a$ can be related by the formula
\begin{equation}
-\lambda\mathcal{F}_{ab}=\zeta_{[a}\sigma_{b]}+\frac{\rm i}{2}\eta_{abcd}\zeta^c\sigma^d. 
\label{eq:decomposeF}
\end{equation}

See \cite{HEUSLER,MARS-EHLERS} for a more complete review of the properties of the quantities just introduced. 
\subsubsection{A local characterisation of the Kerr solution}
We state next a local characterisation of the Kerr solution due to Mars (Theorem 2 of \cite{MARS-EHLERS}).
This local characterisation is formulated exclusively in terms of local properties fulfilled by a Killing vector in a Ricci 
flat spacetime.
\begin{theorem}
Let $(\mathcal{M},g_{ab})$ be a Ricci-flat, non-locally-flat, $C^3$ spacetime possessing a Killing vector $\vzeta$.  
Assume also that $(\mathcal{F}\cdot \mathcal{F})|_p\neq 0$ for at least one point $p\in\mathcal{M}$ and that the condition
\begin{equation}
\mathcal{C}_{abcd}=6L\left(\mathcal{F}_{ab}\mathcal{F}_{cd}-\frac{\mathcal{I}_{abcd}}{3}\mathcal{F}\cdot \mathcal{F}\right) \;,\quad 
{\mathcal I}^{ab}{}_{cd}\equiv\frac{1}{4}(\mbox{i}\;\eta^{ab}{}_{cd}+\delta^a_c\delta^b_d-\delta^a_d\delta^b_c)\;,
\label{eq:kerr-cond-1}
\end{equation}
holds everywhere on $\mathcal{M}$.
Then, the Ernst 1-form  $\sigma_a$ is exact, $L$ vanishes nowhere
and we have:
\begin{equation}
(\mathcal{F}\cdot \mathcal{F})=B(\sigma')^4\;,\quad \sigma'\equiv-\frac{1}{L} \;,
\label{eq:FdotF-property}
\end{equation}
where $\sigma'$ is such that $\nabla_a\sigma'=\sigma_a$ and $B$ is a complex constant. 
If in addition {\em Re}$(\alpha)=$ {\em Re}$(\sigma')-\lambda>0$, 
and $B$ is real and negative then $\mathcal{M}$ is locally isometric to the Kerr
space-time. 
\label{theo:local-kerr-generic}
\end{theorem}
This characterisation is the starting point of our work. Our aim is to write this local characterisation as a single 
scalar condition where the scalar involved has well-defined positivity properties.

\section{The Mars-Simon tensors and their properties}
\label{sec:mars-simon-tensor}
At those points of $\mathcal{M}$ in which the Ernst 1-form is exact and its potential $\sigma'$ does not vanish 
we can introduce the following rank-4 tensor 
\begin{equation}
\mathcal{S}_{abcd}\equiv
\mathcal{C}_{abcd}+\frac{6}{\sigma'}
\left(\mathcal{F}_{ab}\mathcal{F}_{cd}-\frac{\mathcal{I}_{abcd}}{3}\mathcal{F}\cdot \mathcal{F}\right)\;,
\label{def:ms-tensor}
\end{equation}
 Observe that $\mathcal{S}_{abcd}$ is affected by the freedom (\ref{eq:free}) so that one has to either 
(i) try to choose a preferred potential $\sigma'$, or (ii) consider  
families of tensors $\mathcal{S}(\sigma')_{abcd}$ depending on the freedom given by $\alpha$ in (\ref{eq:free}). 
Unless otherwise stated, we adopt in this work the second 
point of view although in general we shall keep the notation $\mathcal{S}_{abcd}$ for the sake of simplicity.

The tensor $\mathcal{S}_{abcd}$ was first introduced in \cite{IONESCU-KLAINERMAN}\footnote{The definition of \cite{IONESCU-KLAINERMAN} 
uses a different choice of the Ernst potential.}
and it was called the {\em Mars-Simon tensor}, a terminology which we also adopt here.
This tensor satisfies a number of remarkable algebraic and differential identities whose proof can be found in 
\cite{IONESCU-KLAINERMAN,GAR-VAL-KERR}. The most relevant ones for us come from the fact that the Mars-Simon tensor is a {\em Weyl field} (also called a Weyl candidate), which means that 
it fulfills the same algebraic properties as the Weyl tensor
\begin{equation}
\mathcal{S}_{abcd}=-\mathcal{S}_{bacd}\;,\quad \mathcal{S}_{abcd}=\mathcal{S}_{cdab}\;,\quad
\mathcal{S}_{abcd}+\mathcal{S}_{acdb}+\mathcal{S}_{adbc}=0\;,\quad
\mathcal{S}^a_{\phantom{a}bac}=0
\label{eq:weyl-candidate}
\end{equation}
and it is self-dual
\begin{equation}
{\rm i}\;\mathcal{S}^*_{abcd}={\rm i}\ ^*\mathcal{S}_{abcd}=\mathcal{S}_{abcd}.
\end{equation}
Its covariant divergence is given in Ricci-flat spacetimes by \cite{IONESCU-KLAINERMAN,GAR-VAL-KERR}
\begin{equation}
\sigma'\nabla_a\mathcal{S}^a_{\phantom{a}bcd}=2\zeta^a(2(\mathcal{F}^p_{\phantom{p}[d}\mathcal{S}_{c]bap}+
\mathcal{F}_b^{\phantom{b}p}\mathcal{S}_{dcap})+{\rm g}_{b[d}\mathcal{S}_{c]apq}\mathcal{F}^{pq}). 
\label{eq:covdiv-ms}
\end{equation}
We note the important property that this covariant divergence is a linear expression
in the Mars-Simon tensor itself.
Using this one can further show that $\mathcal{S}_{abcd}$ fulfills 
a hyperbolic equation \cite{IONESCU-KLAINERMAN} and that it vanishes in a globally hyperbolic 
space-time if it is 
zero on a Cauchy hypersurface (causal propagation) \cite{GAR-VAL-KERR}.

The tensor $\mathcal{S}_{abcd}$ plays an important role in characterizing invariantly the Kerr solution. 
One of such characterizations requires that the space-time be asymptotically flat as originally  
formulated  by Mars in \cite{MARS-KERR}, Theorem {\ref{th:mars} below. To that purpose, 
we briefly introduce the concept of stationary asymptotically flat end.

\subsection{Asymptotically flat end ${\mathcal M}_\infty$ \label{afe}} 
Assume that there exists a submanifold $\mathcal{M}_{\infty}\subset\mathcal{M}$ diffeomorphic to 
$\mathbb{R}\times\{x\in\mathbb{R}^3:|x|>R\}$ for some $R$ large enough. Suppose further, that in the local Cartesian
coordinates $(t,x^i)$, $i=1,2,3$ on $\mathcal{M}_{\infty}$ defined by the diffeomorphism  
we have $\vzeta=\partial_{t}$ (time translation at infinity) and also the conditions
\begin{equation}
|g_{ab}-\eta_{ab}|+|r\partial_{x^i}(g_{ab})|\leq A r^{-B}\;,\quad
\partial_t g_{ab}=0, 
\end{equation}
where $A$ and $B$ are positive constants, $g_{ab}$ and $\eta_{ab}$ represent, respectively, the components 
of the metric and the Minkowski metric in the above local coordinates and $r\equiv\sqrt{(x^1)^2+(x^2)^2+(x^3)^2}$.
A set $\mathcal{M}_{\infty}$ with these properties is called a {\em stationary asymptotically flat end}.
It is well-known (see e.g. \cite{BEIG-SIMON}) that, using the vacuum Einstein equations (without cosmological constant), 
the components of the metric tensor have the following asymptotic behaviour in $\mathcal{M}_{\infty}$ 
\begin{equation}
 g_{tt}=-1+\frac{2M}{r}+O(r^{-2})\;,
\quad g_{tx^i}=-\epsilon_{ijk}\frac{4 S^ix^k}{r^3}+O(r^{-3})\;,\quad
g_{x^ix^j}=\delta_{ij}+O(r^{-1}),
\label{eq:asymptotic-flat-end}
\end{equation}
where $M$ is the Komar mass \cite{KOMAR} of the Killing vector $\vzeta$ in 
$\mathcal{M}_{\infty}$, $\epsilon_{ijk}$ is the Levi-Civita symbol in dimension 3,  and $\delta_{ij}$ is the Kronecker 
symbol (Riemannian flat metric) in dimension 3. 
\subsection{Two theorems on $\mathcal{S}_{abcd}$}
In this subsection we present two results which involve the tensor $\mathcal{S}_{abcd}$
(theorems \ref{th:mars} and \ref{th:kerr-local} below). 
Although not stated explicitly, they are proved in \cite{MARS-KERR}. 

\begin{theorem}
Let $(\mathcal{M},g_{ab})$ be a smooth Ricci-flat spacetime possessing a Killing vector $\vzeta$ and a stationary 
asymptotically flat end ${\mathcal M}_{\infty}$ such
that $\vzeta$ tends to a time translation at infinity and the Komar mass of 
$\vzeta$ at ${\mathcal M}_{\infty}$ does not vanish. Assume further that $\sigma'$ exists globally and
fix the freedom (\ref{eq:free}) 
by letting $\sigma' \rightarrow 0$ at ${\mathcal M}_{\infty}$. Construct the tensor 
$\mathcal{S}_{abcd}$ according to (\ref{def:ms-tensor}).

If $\mathcal{S}_{abcd}$ vanishes, then $(\mathcal{M},g_{ab})$ is locally isometric to the Kerr spacetime. 
\label{th:mars}
\end{theorem}

This theorem includes a global condition, namely, the existence of a stationary asymptotically 
flat end ${\mathcal M}_{\infty}$. This is important to fix the freedom (\ref{eq:free}) 
for the potential $\sigma'$ to be used in the definition of ${\mathcal S}_{abcd}$, 
as one can choose the potential such that $\sigma'$ goes to 0 at the asymptotically flat end 
so that Theorem \ref{th:mars} possesses a totally unambiguous meaning.

Nevertheless, it is possible
to formulate a purely local characterization of the Kerr metric, see Theorem 1 in \cite{MARS-KERR-UNIQUENESS}. 
To that end, we need to control the change of $\mathcal{S}_{abcd}$ under (\ref{eq:free}), and thus we introduce the tensor 
\begin{equation}
\mathcal{Q}_{abcd}\equiv 
6\left(\mathcal{F}_{ab}\mathcal{F}_{cd}-\frac{\mathcal{I}_{abcd}}{3}\mathcal{F}\cdot \mathcal{F}\right).
\label{eq:q-tensor} 
\end{equation}
The algebraic properties of $\mathcal{F}_{ab}$ and $\mathcal{I}_{abcd}$ guarantee that 
$\mathcal{Q}_{abcd}$ is a Weyl candidate too.
It is a very simple matter to check that under the freedom (\ref{eq:free}) the Mars-Simon tensor changes as
\begin{equation}
\mathcal{S}(\sigma')_{abcd}=\mathcal{S}(\sigma)_{abcd}-\frac{\alpha}{\sigma(\sigma +\alpha)} \mathcal{Q}_{abcd}
\label{eq:change}
\end{equation}
Therefore the existence of a potential such that the Mars-Simon tensor va\-ni\-shes is equivalent to the proportionality of 
$\mathcal{S}_{abcd}$ ---constructed from {\em any} given $\sigma'\neq 0$--- 
and $\mathcal{Q}_{abcd}$, with some given properties of the proportionality function. This will be relevant in section 
\ref{sec:qfactors} and allowed Mars to prove his second, purely local, theorem.
\begin{theorem}
Let $(\mathcal{M},g_{ab})$ be a smooth non-flat 
 Ricci-flat spacetime possessing a Killing vector $\vzeta$ and assume further that the tensor 
$\mathcal{S}_{abcd}$ is well defined on $\mathcal{M}$. If there exists a potential $\sigma'$ in (\ref{eq:free}) 
such that the following conditions are fulfilled on $\mathcal{M}$
\begin{equation}
\mathcal{S}(\sigma')_{abcd}=0\;,\quad  \mathcal{F}\cdot\mathcal{F}=-\frac{\sigma'^4}{4M^2}\;,\quad M\in\mathbb{R}\setminus\{0\}\;,
\quad \mbox{\em Re}(\sigma')-\lambda>0\;,
\end{equation}
then $\mathcal{M}$ is locally isometric to the Kerr space-time.
\label{th:kerr-local}
\end{theorem}

\begin{remark}
The constant $M$ here is the Komar mass of $\vzeta$ as introduced in subsection \ref{afe}. 
Note that the imposed condition $M\neq 0$ on the Komar mass excludes the flat Minkowski solution
even though strictly speaking it belongs to the Kerr family. In this work ``Kerr space-time'' 
will mean any member of the Kerr family with non-zero mass parameter.  
\end{remark}

\subsection{The space-time Simon tensor}
The vanishing of $\mathcal{S}(\sigma')_{abcd}$ is of course equivalent to a particular proportionality 
between $\mathcal{C}_{abcd}$ and $\mathcal{Q}_{abcd}$:
\begin{eqnarray}
&&\mathcal{S}(\sigma')_{abcd}=0 \quad \Longleftrightarrow \quad 
\mathcal{C}_{abcd}=-\frac{1}{\sigma'}\mathcal{Q}_{abcd} \quad \Longleftrightarrow\nonumber\\
&&\exists \alpha\in\mathbb{C}: \,\,\, \mathcal{C}_{abcd}=-\frac{1}{\sigma +\alpha}\mathcal{Q}_{abcd}
\label{eq:proportionality}
\end{eqnarray}
and this is why it will be convenient, and very useful, to use another tensor which measures such a proportionality. To that end, let us define
\begin{equation}
\mathcal{S}_{abc}\equiv\gamma_{a[b}\mathcal{C}_{c]mrd}\zeta^m
\mathcal{F}^{rd}+4\zeta^m\zeta^r\mathcal{C}_{mar[c}\sigma_{b]}
\label{eq:simon-tensor}
\end{equation}
where here and in the rest of the paper we use the abbreviation
\begin{equation}
\gamma_{ac}\equiv \zeta_a\zeta_c-\lambda g_{ac} \, .
\label{eq:gamma}
\end{equation}

The tensor $\mathcal{S}_{abc}$ is called the {\em space-time Simon
tensor} 
\cite{MARS-KERR} and it can be defined in any spacetime with a Killing vector, 
independently of the existence of a potential 
$\sigma$ for $\sigma_a$. This will be very important later.
$\mathcal{S}_{abc}$ has the algebraic properties of a Lanczos potential
\begin{equation}
\mathcal{S}^b_{\phantom{a}bc}=0\;,\quad
\mathcal{S}_{a[bc]}=\mathcal{S}_{abc}\;,\quad
\mathcal{S}_{abc}+\mathcal{S}_{bca}+\mathcal{S}_{cab}=0
\label{eq:lanczos}
\end{equation}
hence the space-time Simon tensor is a {\em Lanczos field} (or a Lanczos candidate). From (\ref{eq:simon-tensor}) we deduce the additional 
properties
\begin{equation}
\zeta^a\mathcal{S}_{abc}=0\;,\quad  \zeta^b\mathcal{S}_{abc}=0
\label{eq:spatial-ms}
\end{equation}
so that $\mathcal{S}_{abc}$ is totally {\em orthogonal} to the Killing vector $\vzeta$ from which it is constructed.

A fundamental property of this tensor is (see considerations leading to eq. (16) in \cite{MARS-KERR}).
\begin{lemma}
In arbitrary spacetimes with a non-null Killing vector 
$\vzeta$ such that $\mathcal{F}\cdot \mathcal{F}\neq 0$, 
the vanishing of the space-time Simon tensor 
is equivalent to the proportionality of the self-dual tensors $\mathcal{C}_{abcd}$ and $\mathcal{Q}_{abcd}$:
$$
\mathcal{S}_{abc}=0  \hspace{4mm} \Longleftrightarrow \hspace{4mm} \mathcal{C}_{abcd}= L \mathcal{Q}_{abcd} 
$$
where $L$ is a complex function.
\label{lem:simon=0}
\end{lemma}
As seen above in (\ref{eq:proportionality}), the vanishing of $\mathcal{S}_{abcd}$ is equivalent to 
the proportionality between $\mathcal{C}_{abcd}$ and $\mathcal{Q}_{abcd}$ whenever there exists a potential $\sigma'$ 
for the Ernst one-form ---so that the Mars-Simon tensor can be defined---. Thus, Lemma \ref{lem:simon=0} 
implies the existence of a precise relation between the tensors 
$\mathcal{S}_{abc}$ and $\mathcal{S}_{abcd}$. 
This relation must be invariant under the freedom (\ref{eq:free}), 
because so is $\mathcal{S}_{abc}$, and given that $\mathcal{S}_{abcd}$ changes 
according to (\ref{eq:change}) this relation will not involve $\mathcal{Q}_{abcd}$. 
This leads to exactly the same relation (\ref{eq:simon-tensor}) but simply replacing $\mathcal{C}_{abcd}$ by $\mathcal{S}_{abcd}$:
\begin{equation}
\mathcal{S}_{abc}=\gamma_{a[b}\mathcal{S}_{c]mrd}\zeta^m
\mathcal{F}^{rd}+4\zeta^m\zeta^r\mathcal{S}_{mar[c}\sigma_{b]}
\end{equation} 

The main relevance of the space-time Simon tensor resides in the following result
(lemma 4 in \cite{MARS-KERR}). 
\begin{prop}
Let $(\mathcal{M},g_{ab})$ be a Ricci-flat, non-locally-flat, $C^3$ spacetime possessing a Killing vector 
$\vzeta$ which is non-null on a dense subset of $\mathcal{M}$ and assume
that the corresponding spacetime Simon tensor
vanishes everywhere. Assume also that $\mathcal{F}\cdot \mathcal{F}\neq 0$ everywhere. 
Then, the Ernst 1-form is exact $\sigma_a =\nabla_a \sigma'$ and furthermore:
$$
\mathcal{C}_{abcd}=-\frac{1}{\sigma'}\mathcal{Q}_{abcd}\;,\quad \mathcal{F}\cdot\mathcal{F}= B(\sigma')^4
$$
where $B$ is a complex constant.
\label{prop:simon=0}
\end{prop}
Combining this result with Theorem \ref{theo:local-kerr-generic} 
we obtain another version of the local characterization of the Kerr space-time.
\begin{theorem}
Under the same hypothesis of Proposition \ref{prop:simon=0}, $(\mathcal{M},g_{ab})$ is locally isometric to the Kerr space-time if 
and only if Re($\sigma')-\lambda>0$ and $B$ is real and negative.
\label{th:kerr-local2}
\end{theorem}

\section{The superenergy of $\mathcal{S}_{abcd}$}
\label{sec:superenergy-mars-simon}
 Given a real Weyl candidate $W_{abcd}$, we define its Bel-Robinson
tensor $T_{abcd}\{W\}$ by 
\begin{equation}
T_{abcd}\{W\}\equiv
W_{a\phantom{p}d}^{\phantom{a}p\phantom{d}m}
W_{bp cm}+
W_{a\phantom{\sigma}c}^{\phantom{a}m\phantom{c}p}
W_{bm dp}
-\frac{1}{8}g_{ab}g_{cd}W_{mnpl}
W^{mnpl}.
\label{b-r}
\end{equation}
As described in the general framework of \cite{SUPERENERGY}, 
this tensor is the basic {\em superenergy tensor} of the Weyl candidate $W_{abcd}$ in 4 dimensions. We
prefer to call this tensor the Bel-Robinson tensor of
$W_{abcd}$ in analogy with the Bel-Robinson tensor
constructed out of the Weyl tensor \cite{BEL-RADIATION,SUPERENERGY}. 

In the present context it is more convenient to 
re-write (\ref{b-r}) in terms of the self-dual tensor $\mathcal{W}_{abcd}=W_{abcd}+{\rm i} W^*_{abcd}$ corresponding to the Weyl candidate and its dual. 
In this way one can prove that the tensor (\ref{b-r}) can be rewritten simply as 
(see e.g. \cite{PR-RINDLER-1})
\begin{equation}
T_{abcd}\{{\mathcal W}\}\equiv \mathcal{W}_{a\phantom{p}c}^{\phantom{a}p\phantom{d}m}
\mathcal{\overline{W}}_{bp dm} =T_{abcd}\{W\} . \label{eq:nueva}
\end{equation}
As any other superenergy tensor, the Bel-Robinson tensor
of a Weyl candidate has the following properties ---see
\cite{SUPERENERGY} for detailed proofs.
\begin{theorem}\label{sproperties}
If $T_{abmn}\{W\}$ is the Bel-Robinson tensor of the Weyl candidate
$W_{abmn}$ then
\begin{enumerate}
 \item\label{first} $T_{(abmn)}\{W\}=T_{abmn}\{W\}$, 
$T^{a}_{\phantom{a}amn}\{W\}=0$.
\item\label{second}Generalized dominant property: if $\vec{u}_1$, $\vec{u}_2$, 
$\vec{u}_3$, $\vec{u}_4$ are causal future-directed vectors then $T_{abmn}\{W\}u^{a}_1u^{b}_2u^{m}_3u^{n}_4\geq 0$.
The inequality is strict if $\vec{u}_1$, $\vec{u}_2$, $\vec{u}_3$, $\vec{u}_4$ are timelike.
\item\label{third} $T_{abmn}\{W\}=0\Longleftrightarrow {\mathcal W}_{abmn}=0 \Longleftrightarrow W_{abmn}=0
\Longleftrightarrow\exists$  a set of timelike vectors $\vec{u}_1$, $\vec{u}_2$, $\vec{u}_3$ and $\vec{u}_4$ such that 
$T_{abmn}\{W\}u^{a}_1u^{b}_2u^{m}_3u^{n}_4=0 \Longleftrightarrow$ there is a timelike vector $\vec u$ such 
that $T_{abmn}\{W\}u^{a}u^{b}u^{m}u^{n}=0$.
\end{enumerate}
\label{theo:br-properties}
\end{theorem}

The last point of this theorem enables us to write the vanishing of the 
self-dual Weyl candidate ${\mathcal W}_{abcd}$ as a single scalar 
condition. For an arbitrary timelike vector $\vec{u}$ we define 
the scalar
\begin{equation}
U_{\vec{u}}({\mathcal W})\equiv T_{abmn}\{W\}u^{a}u^{b}u^{m}u^{n}
\label{eq:general-superenergy}
\end{equation}
which, if $W_{abcd}\neq 0$, is a positive quantity according to points
\ref{second} and \ref{third} of Theorem \ref{theo:br-properties}. Therefore one has a natural scalar quantity which enables us 
to measure, for each timelike vector, the proximity to the geometric conditions determined by the tensor condition 
${\mathcal W}_{abcd}=0$. The scalar quantity $U_{\vec{u}}({\mathcal W})$ is called the {\em superenergy density} of $T_{abcd}\{W\}$ with respect to 
$\vec{u}$ \cite{SUPERENERGY}\footnote{The notation differs from that of \cite{SUPERENERGY}.} and it depends in general on the timelike vector $\vec u$. 

However, in those cases where there is a timelike vector defined invariantly one can select a superenergy density that acquires an invariant meaning. We can apply this idea to the particular case in which ${\mathcal W}_{abcd}=\mathcal{S}_{abcd}$. 
At those points of $\mathcal{M}$ in which the Killing vector $\vec{\zeta}$ is timelike the scalar 
$U_{\vec{\zeta}}(\mathcal{S})$ is positive and 
it will vanish if and only if $\mathcal{S}_{abcd}=0$. Given that $\vzeta$ is defined invariantly
we can take the quantity $U_{\vec{\zeta}}(\mathcal{S})$ as a local invariant measure of the deviation of
the spacetime $(\mathcal{M},g_{ab})$ to the geometric conditions entailed by $\mathcal{S}_{abcd}=0$.

\begin{theorem}
\begin{eqnarray}
&& U_{\vec{\zeta}}(\mathcal{S})=\mathcal{E}_{ab}\mathcal{\bar E}^{ab}+
\frac{3}{4|\sigma'|^2}\left(3({\boldsymbol\sigma}\cdot\bar{\boldsymbol\sigma})^2-
|{\boldsymbol\sigma}\cdot{\boldsymbol\sigma}|^2+4\mbox{\em Re}(\sigma'\mathcal{E}_{ab}\bar\sigma^a\bar\sigma^b)\right),
\label{eq:scalar-nonkerrness}
\end{eqnarray}
where
\begin{equation}
\mathcal{E}_{ab}\equiv\mathcal{C}_{apbq}\zeta^p\zeta^q\;,\quad
{\boldsymbol\sigma}\cdot{\boldsymbol\sigma}\equiv\sigma_a\sigma^a\;,\quad
{\boldsymbol\sigma}\cdot\bar{\boldsymbol\sigma}\equiv\sigma_a\bar\sigma^a.
\end{equation}
\label{theo:ms-superenergy}
\end{theorem}

\proof Starting from (\ref{def:ms-tensor}) and using the definition of the Ernst 1-form, a straightforward computation yields
\begin{equation}
\mathcal{U}_{ac}\equiv\mathcal{S}_{abcd}\zeta^b\zeta^d=\mathcal{E}_{ac}+\frac{\mathcal{F}\cdot\mathcal{F}}{2\sigma'}\gamma_{ac}+
\frac{3\sigma_a\sigma_c}{2\sigma'} .
\label{eq:ms-projection} 
\end{equation}
Next, we write the definition of  $U_{\vec{\zeta}}(\mathcal{S})$ adapting the definition of 
$T_{abcd}\{\mathcal{S}\}$ to eq. (\ref{eq:nueva}). The result is 
\begin{equation}
U_{\vec{\zeta}}(\mathcal{S})=\mathcal{U}_{ac}\overline{\mathcal{U }}^{ac}.
\end{equation}
Now we need to replace in this expression the value of $\mathcal{U}_{ab}$ found above and work out the resulting
expression. After a computation, we find
\begin{eqnarray}
&&\mathcal{U}_{ac}\overline{\mathcal{U}}^{ac}=\mathcal{E}_{ab}\mathcal{\bar E}^{ab}+
\frac{3}{2\sigma'}\bar{\mathcal{E}}_{ab} \sigma^{a} \sigma^{b}+
\frac{3}{2\bar\sigma'}\mathcal{E}_{ab}\bar\sigma^{a}\bar\sigma^{b}+\nonumber\\
&&\frac{3}{4|\sigma'|^2}
\bigg(3 (\boldsymbol{\sigma}\cdot\bar{\boldsymbol\sigma})^2 - 
\lambda(\bar{\mathcal{F}}\cdot\bar{\mathcal{F}}{\boldsymbol\sigma}\cdot{\boldsymbol\sigma}+ 
\mathcal{F}\cdot\mathcal{F}\bar{\boldsymbol\sigma}\cdot\bar{\boldsymbol\sigma})
+ \lambda^2 |\mathcal{F}\cdot\mathcal{F}|^2\bigg).
\end{eqnarray}
The final expression (\ref{eq:scalar-nonkerrness}) follows after using in this expression the identity
\begin{equation}
\lambda\, \,  \mathcal{F}\cdot\mathcal{F}={\boldsymbol\sigma}\cdot{\boldsymbol\sigma},
\label{eq:s^2}
\end{equation}
which is a consequence of (\ref{eq:decomposeF}).
\qed

\begin{remark}\em
Note that the result is valid for a Killing vector $\vzeta$ of any causal character
and for any choice of the potential $\sigma'$. 
\end{remark}

One may also use the previous 
definitions and results to introduce the quantity
\begin{equation}
U_{\vzeta}(\mathcal{Q})\equiv T_{abcd}\{\mathcal{Q}\}\zeta^a\zeta^b\zeta^c\zeta^d
\end{equation}
for the Weyl candidate (\ref{eq:q-tensor}).
According to Theorem \ref{theo:br-properties}, the scalar 
$U_{\vzeta}(\mathcal{Q})$ will be non-negative whenever $\vzeta$ is causal. 
Indeed a straightforward computation yields
\begin{equation}
 U_{\vzeta}(\mathcal{Q})=\frac{9({\boldsymbol\sigma}\cdot\bar{\boldsymbol\sigma})^2-
3|{\boldsymbol\sigma}\cdot{\boldsymbol\sigma}|^2}{4}.
\label{eq:UQ}
\end{equation}
Hence (\ref{eq:scalar-nonkerrness}) can be rendered in the form\begin{equation}
 U_{\vzeta}(\mathcal{S})=U_{\vzeta}(\mathcal{C})+\frac{U_{\vzeta}(\mathcal{Q})}{|\sigma'|^2}+
\frac{3\mbox{Re}(\sigma'\mathcal{E}_{ab}\bar\sigma^a\bar\sigma^b)}{|\sigma'|^2}.
\label{eq:non-kerness-short}
\end{equation}

\subsection{Several alternative expressions for the scalar $U_{\vzeta}(\mathcal{S})$}
The superenergy construction can be carried out for arbitrary tensors of any rank, and not only for the Weyl candidates \cite{SUPERENERGY}. For a 2-form $F_{ab}$ the corresponding superenergy tensor $T_{ab}\{F\}$ is simply the standard energy-momentum tensor of a Maxwell field and can be expressed in any one of the following equivalent forms
\begin{equation}
T_{ab}\{F\}\equiv t_{ab} = F_{ac}F_{b}{}^c-\frac{1}{4}g_{ab} F_{cd}F^{cd} = \frac{1}{2} \left(F_{ac}F_{b}{}^c +F^*_{ac}F^*_{b}{}^c \right)= \frac{1}{2} {\mathcal F}_{ac}\overline{{\mathcal F}}_b{}^c \, .
\label{t}
\end{equation}
Using these equations and (\ref{eq:F^2}) we can easily derive the identities 
$$
t_{ab} = t_{ba}\;,\quad
t^a{}_a =0\;,\quad t_{ac}t_b{}^c =\frac{1}{4} g_{ab}\; t_{cd}t^{cd}\;,\quad 
t_{cd}t^{cd} = \frac{1}{16} (\mathcal{F}\cdot \mathcal{F})\ (\bar{\mathcal{F}}\cdot \bar{\mathcal{F}})\;,
$$
and by virtue of the ``Maxwell" equations $\nabla^a {\mathcal F}_{ab}=0$, $t_{ab}$ is divergence free
\begin{equation}
\nabla^a t_{ab}=0\, .
\label{div-free}
\end{equation}
As any other superenergy tensor, it also possesses the dominant property, that is to say, $t_{ab}u_1^au_2^b \geq 0$ for arbitrary future-pointing vectors $\vec u_1$ and $\vec u_2$, and for timelike $\vec u$ one has
$$
t_{ab}u^au^b \geq 0, \hspace{3mm} t_{ab}u^au^b=0 \Longleftrightarrow t_{ab}=0  \Longleftrightarrow F_{ab}=0 \, .
$$

On using Eq.(\ref{eq:decomposeF}) the tensor $t_{ab}$ can be written in our case as
\begin{equation}
t_{ab} = \frac{1}{8\lambda^2}\left[2({\boldsymbol\sigma}\cdot\bar{\boldsymbol\sigma})\zeta_a\zeta_b +\lambda\left(\sigma_a \bar\sigma_b +\bar\sigma_a\sigma_b \right) -\lambda ({\boldsymbol\sigma}\cdot\bar{\boldsymbol\sigma})g_{ab} +2\zeta_{(a}v_{b)}\right]
\label{eq:decomposet}
\end{equation}
where we have introduced the real one-form
$$
v_b \equiv i\eta_{bcde}\bar\sigma^c\zeta^d\sigma^e
$$
which obviously satisfies
$$
v_b\zeta^b =v_b\sigma^b =v_b\bar\sigma^b =0, \hspace{1cm} v_bv^b =\lambda\left[({\boldsymbol\sigma}\cdot{\boldsymbol\sigma})({\bar{\boldsymbol\sigma}}\cdot\bar{\boldsymbol\sigma})-({\boldsymbol\sigma}\cdot\bar{\boldsymbol\sigma})^2 \right] \, .
$$

With these definitions at hand, we can find an interesting expression for ${\mathcal E}_{ac}$. To that end, simply use the first in (\ref{eq:maxwell}) to compute
$$
{\mathcal E}_{ac}=-\zeta^b\nabla_c{\mathcal F}_{ab} = -\frac{1}{2}\nabla_c\sigma_a + 
{\mathcal F}_{ab}\nabla_c\zeta^b =-\frac{1}{2}\nabla_c\sigma_a +\frac{1}{2} {\mathcal F}_{ab}({\mathcal F}_c{}^b+\bar{\mathcal F}_c{}^b)
$$
so that introducing here the identity (\ref{eq:F^2}) together with (\ref{eq:s^2}) and the definition of $t_{ab}$ we arrive at the desired expression
\begin{equation}
{\mathcal E}_{ac}=t_{ac} +\frac{1}{8\lambda} 
({\boldsymbol\sigma}\cdot{\boldsymbol\sigma})g_{ac}-\frac{1}{2} \nabla_c\sigma_a \, .
\label{eq:E}
\end{equation}

Observe that this provides a very simple formula for the electric and magnetic parts of the Weyl tensor with respect 
to the preferred timelike direction $\vec \zeta$ in Ricci-flat stationary spacetimes in terms of the Ernst one-form.

Using (\ref{eq:E}), (\ref{eq:s^2}) and (\ref{eq:ms-projection}) we can also derive
\begin{equation}
\mathcal{U}_{ac}=t_{ac} +\frac{1}{8\lambda} ({\boldsymbol\sigma}\cdot{\boldsymbol\sigma})g_{ac}-\frac{1}{2} 
\nabla_c\sigma_a +\frac{({\boldsymbol\sigma}\cdot{\boldsymbol\sigma})}{2\lambda\sigma'}\gamma_{ac}+\frac{3\sigma_a\sigma_c}{2\sigma'}.
\label{eq:ms-projection2} 
\end{equation}

From the previous formula (\ref{eq:E}) it is very easy to compute the Bel-Robinson superenergy with respect to the direction $\vzeta$:
$$
U_{\vzeta}(\mathcal{C})=\frac{1}{4}\left[\nabla_a\sigma_b\nabla^a\bar\sigma^b -2t_{ab}\nabla^a(\sigma^b+\bar\sigma^b) \right]=
\frac{1}{4}\left[\nabla_a\nabla_b\sigma\nabla^a\nabla^b\bar\sigma-4t_{ab}\nabla^a\nabla^b\lambda \right]\;,
$$
which on using (\ref{eq:decomposet}) can also be written in terms of the Ernst one-form exclusively as
\begin{eqnarray}
&&U_{\vzeta}(\mathcal{C})=\frac{1}{4}\nabla_a\sigma_b\nabla^a\bar\sigma^b -\frac{1}{16\lambda^2}({\boldsymbol\sigma}\cdot{\boldsymbol\sigma})
({\bar{\boldsymbol\sigma}}\cdot\bar{\boldsymbol\sigma})+\frac{({\boldsymbol\sigma}\cdot\bar{\boldsymbol\sigma})}{32\lambda^2}
\left[({\boldsymbol\sigma}\cdot{\boldsymbol\sigma})+({\bar{\boldsymbol\sigma}}\cdot\bar{\boldsymbol\sigma}) \right]\nonumber\\
&&-\frac{1}{8\lambda} \bar\sigma^a\sigma^b\nabla_a(\sigma_b+\bar\sigma_b).
\label{eq:UC}
\end{eqnarray}
This is an expression for the Bel-Robinson super-energy relative to the direction $\vzeta$ involving only the Ernst one-form and its derivatives.

We can also use (\ref{eq:E}) together with (\ref{t}) to compute
$$
{\mathcal E}_{ab}\bar\sigma^a\bar\sigma^b=\frac{({\bar{\boldsymbol\sigma}}\cdot\bar{\boldsymbol\sigma}) }
{8\lambda}\left[({\boldsymbol\sigma}\cdot{\boldsymbol\sigma})+({\boldsymbol\sigma}\cdot\bar{\boldsymbol\sigma}) \right]
-\frac{1}{2}\bar\sigma^a\bar\sigma^b\nabla_b\sigma_a \, .
$$
The previous two expressions together with (\ref{eq:UQ}-\ref{eq:non-kerness-short}) lead after some manipulations to
\begin{eqnarray}
&&U_{\vzeta}(\mathcal{S})=\frac{1}{4}
\nabla_{b}\bar{\sigma}_{a}\nabla^{b}\sigma^{a} +  
\frac{9({\boldsymbol \sigma}\cdot \bar{\boldsymbol \sigma})^2 }
{4|\chi|^2} -(6 \lambda^2-6\lambda +|\chi|^2)
\frac{|{\boldsymbol\sigma}\cdot{\boldsymbol \sigma}|^2}{16\lambda^{2}|
\chi|^2}+\nonumber\\
&&2\mbox{Re}\left[\frac{(\boldsymbol\sigma\cdot\bar{\boldsymbol\sigma})
({\boldsymbol\sigma}\cdot{\boldsymbol\sigma})}{32\lambda^{2}|\chi|^2}
(6 \lambda + \chi) \bar{\chi}-\frac{\sigma^{a} \bar{\sigma }^{b}}
{8\lambda}\nabla_{b}\sigma_{a}-\frac{3\sigma^{a}\sigma^{b}}{4\chi}\nabla_{b}\bar{\sigma}_{a}
\right]\;,
\label{eq:SEdeS}
\end{eqnarray}
where in this expression the choice of the potential $\sigma'=\chi\equiv 1+\lambda+2{\rm i}\omega$ was made 
Note that (\ref{eq:SEdeS}) is an expression depending exclusively on the scalar $\chi$ and its derivatives.

We can summarize the previous results in the following theorem
\begin{theorem}
Let $(\mathcal{M},g_{ab})$ be a smooth spacetime with a Killing vector $\vzeta$ 
that contains a stationary asymptotically flat 4-end ${\mathcal M}_{\infty}$ such that 
$\vzeta$ tends to a time translation at infinity and the Komar mass of $\vzeta$ at ${\mathcal M}_{\infty}$ 
does not vanish. Fix the freedom (\ref{eq:free}) such that $\sigma' \longrightarrow 0$ at ${\mathcal M}_{\infty}$
(this corresponds to the choice $\sigma'=\chi=1+\lambda+2{\rm i}\omega$).

Then, the scalar (\ref{eq:SEdeS}) constructed exclusively from $\chi$ and its first and second derivatives 
is non-negative on the entire region where $\vzeta$ is timelike, and it vanishes if and only if 
$(\mathcal{M},g_{ab})$ is locally isometric to the Kerr spacetime. 
\label{th:U}
\end{theorem}

\section{Quality factors that measure the Kerr-ness of stationary spacetimes}
\label{sec:qfactors}
In this section we present several quality factors to measure how close a given spacetime 
is to the exterior part of the Kerr solution. By a quality factor we mean an adimensional 
scalar function $q$ with $q\in[0,1]$ and such that its value provides a measure of the ``Kerr-ness'' 
of the given space-time at any point.

We present three different possibilities. All of them can be applied to generic stationary spacetimes, 
the first two to stationary Ricci-flat ones, while the first one can only be used when the space-time is 
Ricci-flat and has an asymptotically flat end.

\subsection{A quality factor for stationary Ricci-flat asymptotically flat spacetimes}
Theorem \ref{th:U} provides a neat characterization for the Kerr solution in terms of a single scalar 
and real quantity, which happens to be strictly positive if the metric is not Kerr. 
One can thus think that the scalar (\ref{eq:SEdeS}) may provide a measure of how much 
a region of the spacetime differs from the stationary portion of Kerr spacetime: the smaller the scalar 
(\ref{eq:SEdeS}) is, the better resemblance to the Kerr solution. 
Unfortunately, the previous sentence does not have an absolute meaning, 
because the scalar (\ref{eq:SEdeS}) has physical units ---of (length)$^{-4}$--- 
and thus we can always choose the physical dimensions such that it becomes large or small. 

What is needed here is another similar quantity that normalizes (\ref{eq:SEdeS}) 
and eliminates the physical units. There may be several choices here, but our own, 
which we believe is a natural one, is based again on the superenergy construction 
and its positivity properties. 
As a matter of fact, one can define a (positive-definite and sesquilinear) 
inner product $\left<\, , \, \right>_{\vzeta}$ on the 
complex vector space of self-dual Weyl candidates as follows:
$$
\left<{\mathcal W}_1 , {\mathcal W}_2\right>_{\vzeta}\equiv  \mathcal{W}_1{}_{a\phantom{p}c}^{\phantom{a}p\phantom{d}m}\,  
\mathcal{\overline{W}}_{2\, bp dm}\,  \zeta^a\zeta^b\zeta^c\zeta^d.
$$
Sesquilinearity follows obviously from the definition, while its positive definiteness is a direct consequence 
of Theorem \ref{theo:br-properties}. Observe that the norm here is simply 
the superenergy scalar introduced in (\ref{eq:general-superenergy})
\begin{equation}
\parallel{\mathcal W} \parallel_{\vzeta}^2 \equiv  \left<{\mathcal W} , {\mathcal W}\right>_{\vzeta} = U_{\vzeta}(\mathcal{W}) \, . 
\label{eq:norm}
\end{equation}
\begin{definition}[Quality factor $q$]
The quality factor measuring the {\em Kerr-ness} of any stationary region in a non-flat
asymptotically flat and Ricci-flat spacetime $(\mathcal{M},g_{ab})$ is defined by
$$
q= 1 - Q^2\;,
$$
where 
$$
Q^2 \equiv \frac{U_{\vzeta}(\mathcal{S})}{\left(\sqrt{U_{\vzeta}(\mathcal{C})}+|\chi|^{-1}\sqrt{U_{\vzeta}(\mathcal{Q})}\right)^2}\;,\quad 
\chi=1+\lambda+2{\rm i}\omega.
$$
\label{def:factor-q}
\end{definition}

The main properties of this quality factor are collected in the following
\begin{theorem}
The quality factor $q$ has the following properties:
\begin{enumerate}
\item $q\in \left[0,1\right]$.
\item $q=1$ on an open neighborhood $U\subset \mathcal{M}$ if and only if $(U,g_{ab})$ is locally isometric to the Kerr solution.
\item At any point $x\in \mathcal{M}$, $q=0$ if and only if 
${\mathcal C}_{abcd}|_x = k ( {\mathcal Q}_{abcd}/\chi) |_x$ for a positive real constant $k\in \mathbb{R}^+$.
\end{enumerate}
\label{theo:q-factor}
\end{theorem}
\begin{remark}
{\em Notice that the condition for $q=0$ is that the space-time is of Petrov type D ---as it is the case for the Kerr solution---, however the proportionality constant $k$ has the {\em opposite} sign to the case of Kerr.}
\end{remark}
\proof From (\ref{eq:norm}) we can rewrite $Q^2$ as
$$
Q^2 =\frac{\parallel {\mathcal S}(\chi)\parallel_{\vzeta}^2}{\left(\parallel {\mathcal C}\parallel_{\vzeta}  + 
\parallel\frac{{\mathcal Q}}{\chi}\parallel_{\vzeta}\right)^2}\;,
$$
and using here that (indices suppressed) ${\mathcal S}(\chi)={\mathcal C}+{\mathcal Q}/\chi$ the triangle inequality\\ 
$\parallel {\mathcal C}+{\mathcal Q}/\chi\parallel_{\vzeta}\, \leq \, 
\parallel {\mathcal C}\parallel_{\vzeta}  + \parallel {\mathcal Q}/\chi\parallel_{\vzeta}$ leads to
$$
0\leq Q^2\leq 1\;,
$$
which proves {\em 1}. Now, Theorem \ref{th:U} informs us that the space-time is locally isometric to Kerr if and only 
if $U_{\vzeta}(\mathcal{S}(\chi))=\parallel{\mathcal S(\chi)} \parallel_{\vzeta}^2=0$, that is, if $Q^2=0$, 
which proves {\em 2}. To prove {\em 3}, recall that the triangle inequality is an equality if and only if 
${\mathcal Q}/\chi |_x$ and ${\mathcal C}|_x$ are {\em positively} linearly dependent, i.e. if 
${\mathcal C}|_x = k ({\mathcal Q}/\chi)|_x$ with $k\geq 0$. The case with $k=0$ is excluded as the space-time would be flat. 
\qed

\begin{remark}
It is very important to remark that $q$ can be computed in any given non-flat
Ricci-flat and asymptotically flat stationary space-time {\em without any reference} to the Kerr solution. 
One does not need to map the space-time into the Kerr solution, nor is it necessary to compare or choose coordinates relative 
to Kerr. All that one needs is an explicit expression of the metric of the space-time to be compared to the Kerr solution. 
Notice, furthermore, that using (\ref{eq:SEdeS}),(\ref{eq:UC}) and (\ref{eq:UQ}) one can immediately write an expression for 
$q$ in terms of $\chi$ and its derivatives.
\end{remark}

At any point $x$ of the stationary region in  $(\mathcal{M},g_{ab})$ one can say that $x$ is ``$(q\times 100)$\% Kerr'' (so it is 100\% Kerr if $q=1$, and 0\% Kerr if $q=0$). 
The precise value of $q$ which gives a good approximation to Kerr depends on the context, and it will have to be calibrated 
by studying several examples. Some of these are given in Section \ref{sec:examples}.

\subsection{Quality factors for stationary Ricci-flat spacetimes}

There may arise situations where one does not know if the entire space-time is asymptotically flat, 
or also when one knows that there are no asymptotically flat ends but still the proximity with a Kerr space-time is suspected, 
for example close to the ergo-surface of the Killing vector or to a possible null hypersurface which one suspects may resemble 
the horizon of Kerr space-time. In these situations we 
cannot apply the quality factor introduced in definition \ref{def:factor-q}, which requires fixing the potential 
$\sigma'$ and the existence of an asymptotically flat end.
However, we can still make use of the previous results by complicating things a little bit, as follows. 

For Ricci-flat spacetimes we can make use of Theorems \ref{th:kerr-local} and \ref{th:kerr-local2} which do not require asymptotic flatness. 
The first thing we must require to use those theorems, according to (\ref{eq:proportionality}) or Lemma \ref{lem:simon=0}, 
is the proportionality between $\mathcal{C}_{abcd}$ and $\mathcal{Q}_{abcd}$. Note, however, that this can be easily measured by means of
$$
q_1 \equiv \frac{\left|  \left<\mathcal{C},\mathcal{Q}\right>_{\vzeta}\right|^2 }{\parallel \mathcal{C}\parallel^2_{\vzeta}  \, \,\, \, \,  \parallel \mathcal{Q}\parallel^2_{\vzeta}} \, .
$$
\begin{theorem}
The quality factor $q_1$ has the following properties
\begin{enumerate}
 \item  $q_1\in [0,1]$.
 \item $q_1=1$ at a point $x\in\mathcal{M}$ if and only if $\exists l\in \mathbb{C}: \, \, \mathcal{C}_{abcd}|_x =l \mathcal{Q}_{abcd}|_x$.
 \item $q_1=0$  at a point $x\in\mathcal{M}$ if and only if $\mathcal{E}_{ab}\bar\sigma^a\bar\sigma^b|_x=0$.
\end{enumerate}
\label{theo:q1-factor}
\end{theorem}
\proof 
The Cauchy-Schwarz inequality immediately gives
\begin{equation}
0\leq q_1 \leq 1, \hspace{1cm} q_1|_x =1 \Longleftrightarrow  \mathcal{C}_{abcd} |_x=l \mathcal{Q}_{abcd}|_x ,
\label{eq:q1-properties}
\end{equation}
for some $l\in \mathbb{C}$. The other extreme value, $q_1=0$, can only occur when
$$
\left<\mathcal{C},\mathcal{Q}\right>_{\vzeta}=\frac{3}{2}\mathcal{E}_{ab}\bar\sigma^a\bar\sigma^b\;,
$$
vanishes, proving point 3.
\qed

Point 2 in this theorem informs us that $q_1$ measures 
how much $\mathcal{C}_{abcd}$ and $\mathcal{Q}_{abcd}$ are proportional to each other. 
A possible alternative to measure this proportionality is the use of the space-time Simon 
tensor $\mathcal{S}_{abc}$. We first note the following simple result.
\begin{prop}
Assume that the Killing vector $\vzeta$ is time-like. Then
\begin{equation}
\mathcal{S}_{abc}\overline{\mathcal{S}}^{abc}\geq 0\;,\quad
\mathcal{S}_{abc}\overline{\mathcal{S}}^{abc}=0\Longleftrightarrow \mathcal{S}_{abc}=0. 
\end{equation}
\label{prop:simon-positive}
\end{prop}
\proof Since $\vzeta$ is time-like, from (\ref{eq:spatial-ms}) we know that $\mathcal{S}_{abc}$ 
is totally orthogonal to $\vzeta$, and thus spatial with respect to the Killing vector, 
from where the result follows easily.
\qed

This result together with Lemma \ref{lem:simon=0} 
enables us to use the quantity $\mathcal{S}_{abc}\overline{\mathcal{S}}^{abc}$ 
in a definition of the quality factor to measure the proportionality 
of $\mathcal{C}_{abcd}$ and $\mathcal{Q}_{abcd}$. 
To that end, for any two Lanczos candidates 
$\mathcal{S}^1_{abc}$ and $\mathcal{S}^2_{abc}$ (not necessarily traceless), define
\begin{equation}
\langle\mathcal{S}^1,\mathcal{S}^2\rangle\equiv\mathcal{S}^1_{abc}\overline{\mathcal{S}}^{2abc} .
\end{equation}
This is again a sesquilinear product on the space of Lanczos candidates 
(not necessarily traceless), and it is positive definite when acting on 
candidates that are totally orthogonal to a timelike Killing vector $\vzeta$, 
as follows from Proposition \ref{prop:simon-positive}. Actually, 
$$
\parallel \mathcal{S}\parallel^2 \equiv \left<S,S\right>\;,
$$
happens to be essentially the superenergy of the (real and imaginary parts) of $\mathcal{S}_{abc}$ with respect to $\vzeta$, according to the general construction in \cite{SUPERENERGY}.

Now, observe that $\mathcal{S}_{abc}$ can be written, from its definition 
(\ref{eq:simon-tensor}), directly as the sum $\mathcal{S}^1_{abc}+\mathcal{S}^2_{abc}$ where
$$
\mathcal{S}^1_{abc}\equiv -\frac{1}{2}\gamma_{a[b}\nabla_{c]}\left(\mathcal{F}\cdot \mathcal{F} \right)\;,\quad  \mathcal{S}^2_{abc}\equiv -4\mathcal{E}_{a[b}\sigma_{c]},
$$
which are Lanczos candidates on their own, and they are totally orthogonal to $\vzeta$. 
Therefore we can define 
\begin{equation}
q_2\equiv 1-\frac{\parallel \mathcal{S}\parallel^2}{\left(\parallel \mathcal{S}^1\parallel+\parallel \mathcal{S}^2\parallel \right)^2} \, .
\label{eq:define-q2}
\end{equation}

\begin{theorem}
The quality factor $q_2$ has the following properties 
\begin{enumerate}
 \item  $q_2\in [0,1]$.
 \item $q_2=1$ at a point $x\in\mathcal{M}$ if and only if $\exists l\in \mathbb{C}: \, \, \mathcal{C}_{abcd}|_x =l \mathcal{Q}_{abcd}|_x$.
 \item $q_2=0$  at a point $x\in\mathcal{M}$ if and only if $\nabla_{c}\left(\mathcal{F}\cdot \mathcal{F} \right)|_x=0$ (thus $\mathcal{F}\cdot \mathcal{F}=$const.\ if $q_2=0$ on an open neighborhood) which entails $\mathcal{S}^1_{abc}|_x=0$.
\end{enumerate}
\label{theo:q2-factor}
\end{theorem}
\proof 
The first property is a direct consequence of the triangle inequality \\
$\parallel\mathcal{S}\parallel=\parallel\mathcal{S}^1+\mathcal{S}^2\parallel\leq
\parallel\mathcal{S}^1\parallel+\parallel\mathcal{S}^2\parallel$, while $q_2|_x=1$ happens if and only if 
$\mathcal{S}_{abc}|_x=0$ as is clear from (\ref{eq:define-q2}), so that Lemma \ref{lem:simon=0} entails then 
$\mathcal{C}_{abcd}|_x=l \mathcal{Q}_{abcd}|_x$ for some $l\in\mathbb{C}$. To prove the third point, recall that $q_2|_x=0$ if and only if $\mathcal{S}^1_{abc}|_x$ and $\mathcal{S}^2_{abc}|_x$ are positively proportional to each other, that is 
$$
\mathcal{S}^1_{abc}|_x =k \mathcal{S}^2_{abc}|_x , \hspace{1cm} k\in \mathbb{R}, \, \, \,  k\geq 0
$$
which reads 
$$
 \frac{1}{2}\gamma_{a[b}\nabla_{c]}\left(\mathcal{F}\cdot \mathcal{F} \right)|_x=4k\mathcal{E}_{a[b}\sigma_{c]}|_x \, .
$$
Contracting here $b$ and $a$ we derive
$$
\lambda \nabla_{c}\left(\mathcal{F}\cdot \mathcal{F} \right)|_x =4k \mathcal{E}_{ac}\sigma^a|_x\;,
$$
but using here (\ref{eq:E}) and (\ref{eq:decomposet}) together with (\ref{eq:s^2}) we also have
\begin{equation}
\mathcal{E}_{ac}\sigma^a = 
\frac{{\boldsymbol\sigma}\cdot{\boldsymbol\sigma}}{8\lambda}(\sigma_c +\bar\sigma_c)
-\frac{1}{2}\sigma^b\nabla_c\sigma_b=
\frac{{\boldsymbol\sigma}\cdot{\boldsymbol\sigma}}{4\lambda}\nabla_c\lambda-\frac{1}{4}\nabla_c({\boldsymbol\sigma}
\cdot{\boldsymbol\sigma})=-\frac{\lambda}{4}\nabla_{c}\left(\mathcal{F}\cdot \mathcal{F} \right)\;,
\label{eq:Es}
\end{equation}
so that the previous condition becomes
$$
(1+k)\lambda \nabla_{c}\left(\mathcal{F}\cdot \mathcal{F} \right)|_x=0\;,
$$
which proves the result.
\qed

Actually, eq.(\ref{eq:Es}) allows us to derive an alternative useful expression for $\mathcal{S}_{abc}$ 
(see also \cite{BJM}).
\begin{lemma}
The space-time Simon tensor can be written in the form
\begin{equation} 
\mathcal{S}_{abc}=\frac{1}{\lambda}(4\mathcal{E}_{a[b}\gamma_{c]d}+2\mathcal{E}_{d[c}\gamma_{b]a})\sigma^d. 
\label{eq:simon-tensor-sp}
\end{equation}
\end{lemma}
\proof 
By noting that $\gamma_{cd}\sigma^d =-\lambda\sigma_c$ the first 
summand provides directly $\mathcal{S}^2_{abc}$ while the second summand is $\mathcal{S}^1_{abc}$ as follows from (\ref{eq:Es}).
\qed

If we combine eq. (\ref{eq:q1-properties}) with the second point of Theorem \ref{theo:q2-factor} we deduce that
\begin{equation}
q_1=1 \Longleftrightarrow q_2 =1 \, .
\label{eq:q2q1}
\end{equation}
One can find an alternative, useful, expression for $q_{2}$ as follows. By explicit computation one can easily obtain the following results
\begin{eqnarray}
\parallel \mathcal{S}^1\parallel^2 &=&\frac{\lambda^2}{4} \nabla_c\left(\mathcal{F}\cdot \mathcal{F} \right)
\nabla^c\left(\overline{\mathcal{F}}\cdot \overline{\mathcal{F}} \right)\;, \label{eq:S1^2}\\
\left<\mathcal{S}^1,\mathcal{S}^2\right> &=&-\parallel \mathcal{S}^1\parallel^2\;, \label{eq:S1S2}\\
\parallel \mathcal{S}^2\parallel^2 &=& 8\left[({\boldsymbol\sigma}\cdot\bar{\boldsymbol\sigma})
\mathcal{E}_{ab} \bar{\mathcal{E}}^{ab}- \mathcal{E}_{b}{}^{c} \bar{\mathcal{E}}_{ac} \sigma^{a}\bar{\sigma }^{b} \right]\;,
\label{eq:S2^2}
\end{eqnarray}
where we have used (\ref{eq:Es}) and (\ref{eq:S1^2}) to get (\ref{eq:S1S2}). From these one immediately finds
\begin{equation}
\left<\mathcal{S}^1,\mathcal{S}\right>=0, \hspace{1cm} 
\parallel \mathcal{S}\parallel^2=\parallel \mathcal{S}^2\parallel^2-\parallel \mathcal{S}^1\parallel^2\;, \label{eq:S^2}
\end{equation}
which also implies
$$
\parallel \mathcal{S}^2\parallel^2\geq \parallel \mathcal{S}^1\parallel^2 \, .
$$
Observe that the second in (\ref{eq:S^2}) allows us to rewrite $q_2$ as given in (\ref{eq:define-q2}) in a simpler form
$$
q_2= 1-\frac{\parallel \mathcal{S}^2\parallel^2-\parallel \mathcal{S}^1\parallel^2}{\left(\parallel 
\mathcal{S}^1\parallel+\parallel \mathcal{S}^2\parallel \right)^2} =1-\frac{\parallel \mathcal{S}^2\parallel-
\parallel \mathcal{S}^1\parallel}{\parallel \mathcal{S}^1\parallel+\parallel \mathcal{S}^2\parallel }=
\frac{2\parallel \mathcal{S}^1\parallel}{\parallel \mathcal{S}^1\parallel+\parallel \mathcal{S}^2\parallel }\;,
$$
from where the properties in Theorem \ref{theo:q2-factor} become more transparent.

\begin{remark}
We note that the quality factors $q_1$ and $q_2$ do not involve any potential $\sigma'$ and hence 
they can be computed for any space-time possessing a Killing vector, be it Ricci flat or not.
Also, if we multiply the Killing vector $\vzeta$ by a real constant $k$ then the quality factors $q_1$, $q_2$
computed with respect to $k\vzeta$ remain the same (they are scale invariant).  
\label{rem:q1q2}
\end{remark}

Assume then that any of $q_1$ or $q_2$ were unity. 
In that case, from Proposition \ref{prop:simon=0} we know that, 
for Ricci-flat spacetimes, there would exist a complex potential $\sigma'$ for the Ernst one-form such 
that $\mathcal{C}_{abcd}=(-1/\sigma') \mathcal{Q}_{abcd}$.
Then we would have
\begin{equation}
\left<\mathcal{C},\mathcal{Q}\right>_{\vzeta} =-\frac{1}{\sigma'} \parallel \mathcal{Q}\parallel^2_{\vzeta}\;,
\label{eq:s+a}
\end{equation}
so that the Kerr condition Re$(\sigma')-\lambda>0$ (see Theorem \ref{th:kerr-local2}) would require that
$$
\beta+\lambda<0\;,\quad \beta \equiv \mbox{Re}\left(\frac{\parallel\mathcal{Q}\parallel_{\vzeta}^2}{\left<\mathcal{C},\mathcal{Q} \right>_{\vzeta}} \right).
$$
Notice once again that to compute $\beta$ we only need to know the space-time and the Killing vector, 
and therefore we can control 
this property by means of the quality factor
\begin{equation}
\kappa\equiv 1-\frac{(\beta+\lambda+|\beta+\lambda|)^2}{4|\beta+\lambda|^2}=
\left\{\begin{array}{c}
        +1\;,\quad \beta+\lambda<0\\
         0 \;,\quad \beta+\lambda>0
       \end{array}
\right.
\label{eq:define-kappa}
\end{equation}

\vspace{2mm}
Provided that $q_1=1$ (or equivalently $q_2=1$) we must finally control the second condition on Theorem  \ref{th:kerr-local2}, 
that is to say that the constant $B$ in Proposition \ref{prop:simon=0} be real and negative. From (\ref{eq:s+a}) and (\ref{eq:s^2}) 
this can be rewritten as 
$$
z\equiv ({\boldsymbol\sigma}\cdot{\boldsymbol\sigma})
\frac{\left<\mathcal{C},\mathcal{Q}\right>_{\vzeta}^4 }{\parallel\mathcal{Q}\parallel^8_{\vzeta}}\;,  
$$
being real and positive (as this would have to be $z=B\lambda$). Then set
$$
\varkappa \equiv 1-\frac{1}{4} \left(\frac{z}{|z|}-1\right) \left(\frac{\bar{z}}{|z|}-1\right) \, .
$$

\begin{prop}
The following statements hold true
\begin{itemize}
\item $0\leq \varkappa \leq 1$,
\item $\varkappa =0$ if and only if $z=-|z|=\bar z$ (i.e. $z$ is real and {\em negative}).
\item $\varkappa =1/2$ if and only if $z=-\bar z$ (i.e. $z$ is purely imaginary).
\item $\varkappa =1$ if and only if $z=|z|=\bar z$ (i.e. $z$ is real and {\em positive}). This is the Kerr condition.
\item $\varkappa$ is invariant under the change $\vzeta\rightarrow k\vzeta$, $k\in\mathbb{R}$.
\end{itemize}
\label{prop:varkappa}
\end{prop}
\proof The definition of $\varkappa$ in terms of $z$, $\bar z$ and $|z|$ yields directly the fourth statement.
Also if we set $z/|z|=e^{\rm{i}\phi}$ then elementary manipulations show that 
the scalar $\varkappa$ takes the following form in terms of $\phi$
\begin{equation}
\varkappa=\cos^2\left(\frac{\phi}{2}\right), 
\end{equation}
which makes it clear that $0\leq\varkappa\leq 1$. Now, one has 
\begin{itemize}
 \item $\varkappa=0\Longleftrightarrow\phi=(2n-1)\pi$, $n\in\mathbb{Z}\Longleftrightarrow$  $z$ is real and negative.
 \item $\varkappa=1/2\Longleftrightarrow\phi=\pi/2+n\pi$, $n\in\mathbb{Z}\Longleftrightarrow$ $z$ is purely imaginary.
 \item $\varkappa=1\Longleftrightarrow\phi=2n\pi$, $n\in\mathbb{Z}\Longleftrightarrow$ $z$ is real and positive.
\end{itemize}
\qed

Thus, we can define two new quality factors to measure the Kerr-ness of general stationary Ricci-flat spacetimes.

\begin{definition}[Quality factors $\hat q$ and $\tilde q$]
Two quality factors measuring the {\em Kerr-ness} of any stationary region in a Ricci-flat spacetime 
$(\mathcal{M},g_{ab})$ are defined by
$$
\hat q\equiv  q_2\kappa\varkappa\;,\quad  \tilde{q} \equiv  q_1\kappa\varkappa\;, 
$$
\end{definition}

The main properties of these quality factors are
\begin{theorem}
The quality factors $\hat q$ and $\tilde q$ have the following properties:
\begin{enumerate}
\item $\hat q$, $\tilde q\in \left[0,1\right]$.
\item On an open neighborhood $U\subset \mathcal{M}$ we have that $\hat q=1\Longleftrightarrow\tilde q =1\Longleftrightarrow(U,g_{ab})$ is locally isometric to the Kerr solution.
\item At any point $x\in \mathcal{M}$, $\hat q|_x=0$ if and only if either $(\beta +\lambda)|_x >0$, or $z|_x$ is real and negative, or $\nabla_{c}\left(\mathcal{F}\cdot \mathcal{F} \right)|_x =0$.
\item At any point $x\in \mathcal{M}$, $\tilde  q=0$ if and only if either $(\beta +\lambda)|_x >0$, 
or $z|_x$ is real and negative, or  $\mathcal{E}_{ab}\bar\sigma^a\bar\sigma^b|_x=0$.
\item $\hat q$ and $\tilde q$ are invariant under the change $\vzeta\rightarrow k\vzeta$, $k\in\mathbb{R}$.
\end{enumerate}
\end{theorem}

\proof
The first item follows from the fact that each quality factor is the product of quantities in the interval $[0,1]$. 
To prove the second item, we note that if $\hat q=1$, then clearly $q_2=1$, $\kappa=1$ and $\varkappa=1$, as $q_2,\kappa, \varkappa\in [0,1]$. 
This together with (\ref{eq:q2q1}) implies that  
$$
\hat q=1 \Longleftrightarrow \tilde q=1.
$$ 
According to Proposition \ref{prop:varkappa}, eq. (\ref{eq:define-kappa}), Theorem \ref{theo:q1-factor} and
Theorem \ref{theo:q2-factor}, the fact that $q_2=1$, $q_1=1$, $\kappa=1$, $\varkappa=1$ 
in an open neighborhood $U\subset \mathcal{M}$ is equivalent to
the conditions of Theorem \ref{th:kerr-local2} being met and hence $(U,g_{ab})$ is locally isometric to the Kerr solution.
Similarly, points 3 and 4 of the theorem follow directly from the mentioned propositions and theorems.
Finally point 5 follows from the fact that $\hat q$ and $\tilde q$ are given in terms of products whose
factors are already invariant under the change $\vzeta\rightarrow k\vzeta$.
\qed

\begin{remark}
One could also have defined $\hat q'\equiv (q_2 +\kappa +\varkappa)/3$ ---and 
the same for $\tilde q$--- which also has properties 1 and 2 of the theorem. However, 
observe that the chosen $\hat q$ is more demanding, in the sense that larger values of 
$\hat q$ are required to have a good resemblance to the Kerr space-time. For example, 
if $q_2=1$, $\kappa =1$ and $\varkappa =1/2$  we would have $\hat q'=5/6$ while $\hat q =1/2$.
\end{remark}

\subsection{How to apply the quality factors to general stationary spacetimes}
The results described so far assume that $(\mathcal{M},g_{ab})$ is a
vacuum solution of the Einstein field equations (without cosmological constant $\Lambda$)
and hence they cannot be used to establish the proximity of a non-Ricci-flat
solution to the Kerr spacetime. 
The problems are of two types. First of all, the Mars-Simon tensor can only be defined on
spacetimes with a potential $\sigma'$ for the Ernst one-form, and this only happens when $\zeta_{[b}R_{a]c}\zeta^c=0$. 
A way to circumvent this is to replace the
Mars-Simon tensor with another quantity 
which can be defined in any spacetime, independently of the existence of the Ernst potential. 
Obviously, the space-time Simon tensor is an useful alternative, as it can be defined in any spacetime with a Killing vector, 
be it Ricci-flat or not, and independently of the existence of a potential $\sigma'$. 
And this is where the second difficulty arises, because the vanishing of the space-time Simon 
tensor implies the proportionality of $\mathcal{C}_{abcd}$ and $\mathcal{Q}_{abcd}$, but the explicit 
form of the proportionality factor given in Proposition \ref{prop:simon=0}, as well as the explicit expression 
appearing there for $\mathcal{F}\cdot \mathcal{F} $, depend crucially on the assumption that the Ricci tensor vanishes.

Nevertheless, it is not only desirable, but actually important, to be able to measure the proximity 
to Kerr for stationary spacetimes with non-vanishing Ricci tensor. And this is important at least in three respects: 
(i) to consider approximate solutions to the Einstein vacuum field equations and check their resemblance to 
the Kerr space-time, (ii) for cases with a cosmological constant $\Lambda$, and (iii) for more general cases such as solutions 
for a perfect-fluid energy-momentum tensor, or matched solutions containing a vacuum region and an interior part, and so on. 
In all cases we are going to argue that we can actually use some of the previous results conveniently adapted 
to the situation under consideration, and with some precautions. 

We split this analysis in two, the case with a potential for the Ernst one-form, and the case without it.

\subsubsection{The case with a potential: $R_{ab}\zeta^b=\mu \zeta_a$} 
This case includes, obviously, the vacuum case with a cosmological constant $R_{ab}=\Lambda g_{ab}$, 
as well as for instance any space-time with a {\em rigidly rotating} fluid, which are characterized 
by having an energy-momentum tensor (ergo a Ricci tensor) with a timelike eigendirection proportional 
to a timelike Killing vector $\zeta$.

Now the relations of Proposition \ref{prop:simon=0} no longer hold in general, as they are derived 
using the Ricci-flat condition. Nevertheless, the factors $\kappa$ and $\varkappa$ can still be defined while 
$q_1$ and $q_2$ still measure the proportionality between $\mathcal{C}_{abcd}$ and $\mathcal{Q}_{abcd}$. 
One can expect intuitively that, provided the space-time is actually close to the Kerr solution somewhere, 
the deviation from Ricci-flatness will be small there, and the relations of Proposition \ref{prop:simon=0} 
will hold {\em approximately}, that is, with some small corrections proportional to the non-zero Ricci components. 
Thus, it seems clear that the quality factors $\tilde q$ and $\hat q$ will still measure the Kerr quality of the space-time under analysis. 
It may also happen that the space-time is asymptotically flat, in which case we can also use $q$. 
In all cases it should be kept in mind, however, that values of $q$, $\tilde q$ and $\hat q$ close to 1 will be relevant only 
in combination with the smallness of these Ricci terms ---compared with some characteristic length squared---.

An obvious example to test the previous arguments is the Kerr-de Sitter space-time. Another important example 
is the rigidly rotating perfect-fluid solution due to Wahlquist, which happens to contain the Kerr spacetime as a particular case. 
These spacetimes are discussed in the last section where the quality factors are explicitly computed and plotted.

\subsubsection{The case without a potential: $\zeta_{[b}R_{a]c}\zeta^c\neq 0$}
This is relevant for example for approximate solutions of the Ricci-flat equations, where despite 
the fact that the Ricci components are assumed to be small one has no extra control on them. 
It is also relevant for more general situations.

In this case there arise some doubts about the factors $\kappa$ and $\varkappa$. 
One possibility to provide them with a reasonable meaning is to also require that 
the deviations from having an Ernst potential is very small. To than end, recall that in a general space-time 
the Ernst one-form satisfies \cite{MARS-EHLERS}
$$
\nabla_{[a}\sigma_{b]}={\rm i} \eta_{abcd}\zeta^c R^d{}_e\zeta^e
$$
which in particular implies that $\nabla_{[a}\sigma_{b]}$ is totally orthogonal to $\zeta^b$. 
Thus, the quantity $\nabla_{[a}\sigma_{b]}\nabla^{[a}\bar\sigma^{b]}$ is non-negative and vanishes 
if and only if so does the exterior differential of $\sigma_a$. We can thus use the following scalar 
$$
q_\sigma \equiv 1-\frac{\lambda^2 \nabla_{[a}\sigma_{b]}\nabla^{[a}\bar\sigma^{b]}}{\gamma^{ac}\gamma^{bd}\nabla_a\sigma_b \nabla_c \bar\sigma_d}
$$
as an appropriate quality factor measuring the vanishing of $\nabla_{[a}\sigma_{b]}\nabla^{[a}\bar\sigma^{b]}$, 
and thereby the local existence of an Ernst potential. Observe that $q_\sigma=1$ if and only if there exists $\sigma'$ 
such that $\sigma_a=\nabla_a\sigma'$ locally and that $q_\sigma$ is invariant under the transformation 
$\vzeta\rightarrow k\vzeta$, $k\in\mathbb{R}$.
Then, one can use either of the following quality factors for general stationary spacetimes
$$
\tilde q q_\sigma, \hspace{1cm} \hat q q_\sigma
$$
keeping in mind that only values of $q_\sigma$ near one provide a reliable result. 

To test the behavior of this refined quality factors we have used the standard Lense-Thirring 
approximate solution for a rotating stationary isolated body and the Kerr-Newman solution of the Einstein-Maxwell equations, 
see Section \ref{sec:examples}.

\section{Examples}
\label{sec:examples}
In this section we present explicit numerical studies of the quality factors introduced above.
\subsection{The Curzon-Chazy solution}
The Curzon-Chazy solution is a vacuum solution to the Einstein equations which is asymptotically flat. It is the simplest
member of the {\em Weyl class} (see \cite{EXACTSOLUTIONS} for further details). The explicit form of the metric is
\begin{equation}
ds^2=-e^{-\frac{2M}{R}}dT^2+e^{\frac{2M}{R}}\left(e^{-\frac{M^2\sin^2\theta}{R^2}}(dR^2+R^2d\theta^2)+R^2\sin^2\theta d\phi^2\right)\;,
\quad M\in\mathbb{R} 
\end{equation}
The metric functions are everywhere regular except at $R=0$. The structure of this singularity is rather complicated, being a directional singularity, 
and it has been studied by a number of authors (see \cite{NEWEXACTSOLUTIONS} and references therein for a good account of this). In the present study we restrict 
our attention to the region $-\infty<T<\infty$, $0<R$, $0<\theta<\pi$ and $0<\phi<2\pi$. In this region there exists an asymptotically 
flat end $\mathcal{M}_{\infty}$ located at $\{R>a\}$, with $a\in\mathbb{R}$ big enough and $\partial/\partial T$ as the asymptotic time 
translation.  Therefore we can use the quality factor $q$
introduced in Definition \ref{def:factor-q} to compute the proximity of this solution to the Kerr space-time. In fact the explicit
form of $q$ for this solution is 
\begin{eqnarray*}
&&q=1-\\
&&\frac{1}{\left(\left(e^{\frac{2 M}{R}}-1\right)\sqrt{M^2 \sin^2\theta\left(M^2-3 M R+3 R^2\right)+3 R^2(M-R)^2}+2\sqrt{3}M R\right)^2}\times\\
&&\left(M^2 \sin^2\theta\left(e^{\frac{2M}{R}}-1\right)^2 \left(M^2+3 R^2\right)-3 M^3 R\sin^2\theta\left(e^{\frac{4M}{R}}-1\right)+\right.\\
&&\left.3 R^2\left(e^{\frac{2 M}{R}}(M-R)+M+R\right)^2\right)\;, 
\end{eqnarray*}
where $\vzeta=\partial/\partial T$. 
In figure \ref{fig:curzon} we present a graph of this quantity for a given value of the parameter $M$.
Observe that $q$ reduces at the axis of symmetry ($\theta=0$) to
$$
q|_{\theta=0}=1-\frac{\left(e^{\frac{2M}{R}}(M-R)+M+R \right)^{2}}{\left(e^{\frac{2M}{R}}|M-R|-|M-R|+2M \right)^{2}}
$$
so that $q|_{\theta=0}=0$ at the axis for $R\leq M$, but $q|_{\theta=0}\neq 0$ at the axis for $R>M$. 
In addition, $q\rightarrow 1$ when $R\rightarrow \infty$.

\begin{figure}[t]
\begin{center}
\includegraphics[width=.6\textwidth]{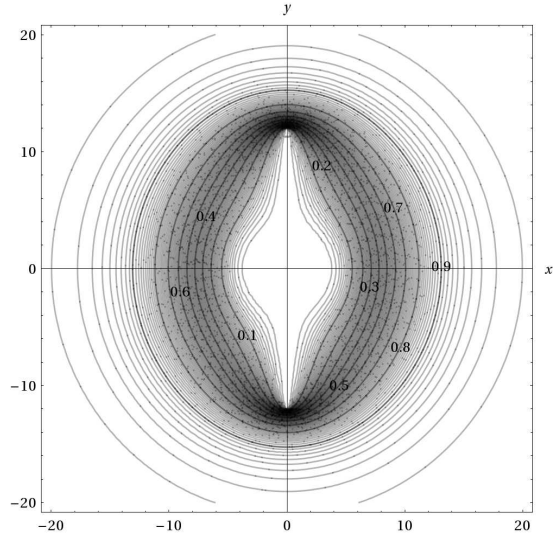}
\caption{\label{fig:curzon} Graphical representation of the quality factor $q$ for the Curzon-Chazy solution when $M=12$. In this picture 
we have defined the variables $y=R\cos\theta$, $x=R\sin\theta$ so the vertical line $x=0$ is the axis of the 
axial Killing vector $\partial/\partial\phi$. The curves are contour lines with constant $q$ and the contour interval is $1/100$. 
We can indeed obtain a contour surface by rotating each contour curve around the axial Killing vector axis. The white central region is 
a set of points where $q<1/100$; actually, $q=0$ on the axis for $R\leq M(=12)$. On the other hand, the outer white region is a set of points where $q>99/100$; in fact, $q\rightarrow 1$ when $R\rightarrow \infty$. The plot indicates the presence of points 
within the axial Killing vector axis where $q$ is not differentiable. These are probably due to the use of square roots in the denominator of $q$, nevertheless $q$ is perfectly continuous everywhere.}
\end{center}
\end{figure}

\subsection{The  $\delta=2$ Tomimatsu-Sato solution}
The Tomimatsu-Sato space-time \cite{TOMIMATSUSATO} can be described by a metric
whose non-vanishing components are given by
\begin{eqnarray}
&& g_{tt}=-\frac{A(x,y)}{B(x,y)}\;,\quad
g_{t\phi} = -\frac{2 M q C(x, y) (-1 + y^2)}{B(x, y)}\;,\quad
g_{xx} = \frac{M^2B(x, y)}{p^{2}\delta^{2}(-1 + x^2)(x^2 -y^2)^{3}}\;,\nonumber\\
&& g_{yy} = \frac{M^2 B(x, y)}{p^{2}\delta^{2}(-1 + y^2)(- x^2 +y^2)^{3}}\;,\nonumber\\
&& g_{\phi\phi} = - \frac{M^2(-1 + y^2)(p^2 B^2(x,y)(-1 + x^2)+
4 q^2 \delta^2 C^2(x,y)(-1 + y^2))}{A(x,y)B(x, y)\delta^{2}}\;,
\end{eqnarray}
$M$, $p$, $q$, $\delta$ being constants, and 
\begin{eqnarray*}
&&A(x,y)=\left(p^2 \left(x^2-1\right)^2+q^2
   \left(1-y^2\right)^2\right)^2-4 p^2 q^2
   \left(x^2-1\right) \left(1-y^2\right)
   \left(x^2-y^2\right)^2,\\
&&B(x,y)=\left(p^2 x^4+2 p x^3-2 p x+q^2 y^4-1\right)^2+4
   q^2 y^2 \left(p x^3-p x y^2-y^2+1\right)^2,\\
&& C(x,y)=p^3 x\left(1-x^2\right) \left(2
   \left(x^4-1\right)+\left(x^2+3\right)
   \left(1-y^2\right)\right)+\nonumber\\
&&p^2\left(x^2-1\right) \left(\left(x^2-1\right)
   \left(1-y^2\right)-4 x^2
   \left(x^2-y^2\right)\right)+q^2
   \left(1-y^2\right)^3 (p x+1).
\end{eqnarray*}
This is a vacuum solution of the Einstein equations if $\delta=2$ and $p^2+q^2=1$. The coordinate ranges 
are $-\infty<t<\infty$, $-1<y<1$, $-\infty<x<-1$, $1<x<\infty$. In the figure \ref{fig:t-s} a numerical study of 
the quality factor $q$ for $\vzeta=\partial/\partial t$ is shown.

\begin{figure*}
\includegraphics[width=0.75\textwidth]{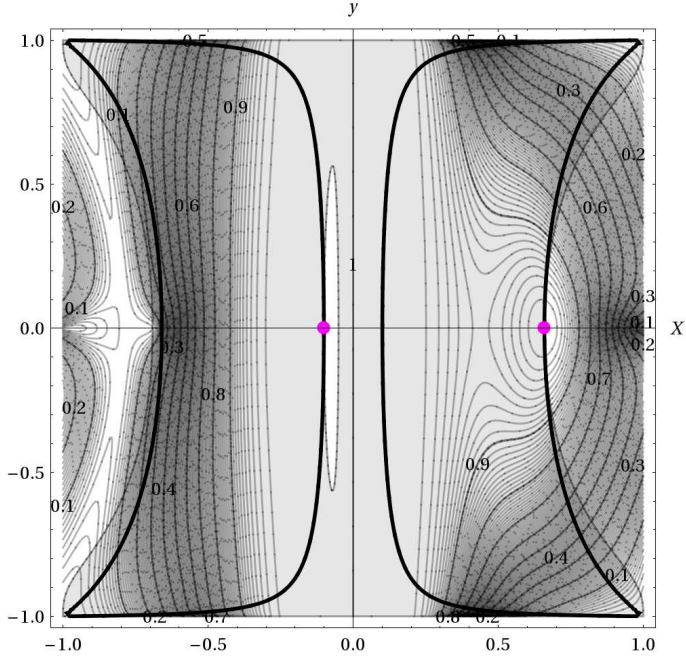}
\caption{\label{fig:t-s}Representation of the Tomimatsu-Sato global structure for $p=1/5$.
The grey region corresponds to those points in which the axial Killing vector 
$\partial/\partial\phi$ is space-like, so no causality violations are expected there. 
The variable $X$ is defined by the condition $X=-1/x$. The lines $X=\pm 1$ correspond to 
conical singularities and the lines $y=\pm 1$ are rotation axes for the Killing vector 
$\partial/\partial\phi$. The vertical line $X=0$ is space-like infinity for two 
different asymptotically flat regions (these are each characterised by $-1<X<0$ and $0<X<1$).
The two black thick closed curves are the ergosurfaces and the two dots are metric (naked) singularities. 
We have superimposed some contour curves for the quality factor $q$ indicating the corresponding value of
$q$ on each curve. In this particular case this scalar turns out to be positive everywhere (and it approaches $q_{\infty}=1$ at $X=0$).
}
\end{figure*}

\subsection{The Wahlquist perfect-fluid space-time}
A rigidly rotating perfect fluid solution to Einstein field equations was found by Wahlquist in \cite{WAHLQUIST}, 
see also \cite{WAHLQUISTESCORIAL,SENOVILLAESCORIAL}, which is known to have a vanishing Simon tensor 
\cite{KRAMERSIMONTENSOR,SENOVILLAESCORIAL} and is the unique stationary and axially symmetric 
such solution of Petrov type D with a particular linear equation of state \cite{SENOVILLAAXISSYMMETRICPF}.
The Wahlquist spacetime actually contains the Kerr-de Sitter space-time as a particular limit case 
\cite{MARSWAHLQUISTNEWMAN}. 
There are several equivalent but different forms for the Wahlquist solution in the literature, 
see \cite{MARSSENOVILLAWAHLQUIST,RACZZSIGRAI} for a discussion of this, 
and the best choice for our purposes is the following in local coordinates $\{t,\varphi,y,z\}$
$$
ds^2 =-\frac{V-U}{v_1+v_2}\left(dt - \frac{v_1V+v_2U}{V-U}d\varphi \right)^2+\frac{UV(v_1+v_2)}{V-U} d\varphi^2+(v_1+v_2)\left(\frac{dy^2}{V} + \frac{dz^2}{U} \right)
$$
where $v_1(z)=(\cosh(2\beta z)-1)/(2\beta^2)$, $v_2(y)=(1-\cos(2\beta y))/(2\beta^2)$, and 
\begin{eqnarray*}
U(z)= Q_0-\left(\nu_0+\frac{\mu_0}{\beta^2}\right) v_1+\frac{\sinh(2\beta z)}{2\beta} \left(a_1+z\frac{\mu_0}{\beta^2} \right)\\
V(y)= Q_0+\left(\nu_0+\frac{\mu_0}{\beta^2}\right) v_2+\frac{\sin(2\beta y)}{2\beta} \left(a_2-y\frac{\mu_0}{\beta^2} \right)
\end{eqnarray*}
with $Q_0,a_1,a_2,\nu_0,\mu_0$ and $\beta$ arbitrary constants. The Kerr-de Sitter-NUT line element is obtained by setting 
$\beta =0$, and $a_1$ is essentially the NUT parameter. Thus, good values of the quality factors are only expected for small $\beta$ and $a_1$.
For this solution, the space-time Simon tensor vanishes identically and therefore the quantity $q_1$ (and hence $q_2$) is identically 1. 
Also the Ernst 1-form is closed so there is a local Ernst potential (or equivalently $q_{\sigma}=1$).
This means that $\hat{q}q_{\sigma}=\tilde{q}q_{\sigma}=\kappa\varkappa$. We provide a numerical study of $\kappa\varkappa$
computed for the Killing vector $\vzeta=\partial/\partial t$ in figure
\ref{fig:wahlquist}. 
\begin{figure}[t]
\includegraphics[width=.5\textwidth]{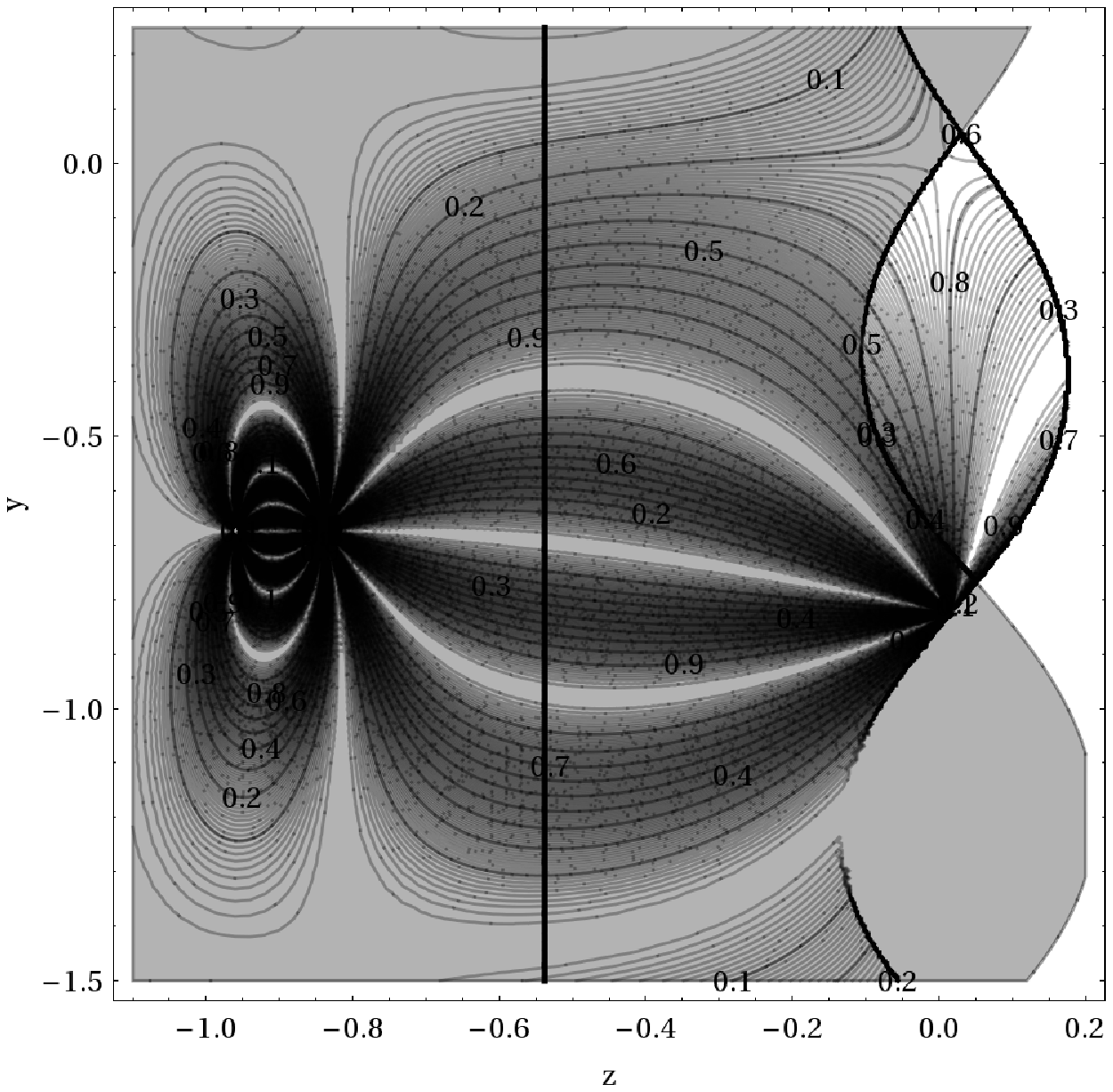}
\includegraphics[width=.5\textwidth]{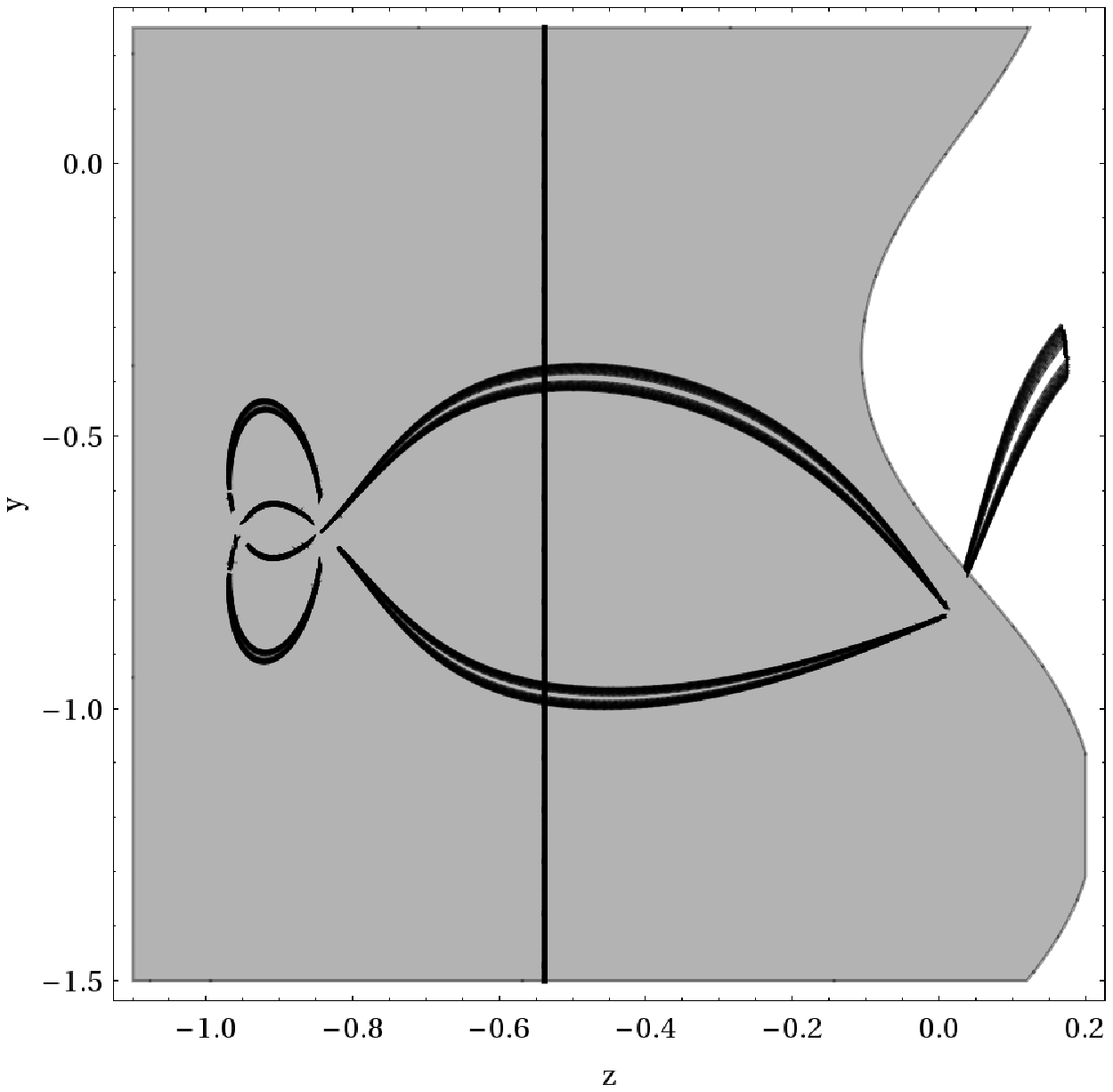}
\caption{\label{fig:wahlquist} 
In these figures we show a representation of the quality factor $\hat q =\tilde q= \kappa\varkappa$ 
for the Wahlquist solution with parameters $Q_0=4$, $a_1=3.015$, $a_2=1.585$, $\nu_0=0.64$, $\mu_0=1.265$, 
$\beta=1.88$ and $\vzeta=\partial/\partial t$. The picture on the left is a representation of the factor 
$\kappa\varkappa$ in the plane $z-y$ by contour curves (contour interval 1/100). To facilitate the understanding of this graph, 
the coordinate $z$ has been compactified 
by means of the transformation $z\rightarrow \tan z$ and the grey region is the region where $\vzeta$ is causal. The black 
vertical line corresponds to points where the determinant spanned by the Killing vector vanishes. The picture on the right
is the same as the picture on the left but only the the region where $\hat q>99/100$ is represented (in black).
The pattern shown in these pictures repeates itself {\em quasi-periodically} along the y-axis.}
\end{figure}

\subsubsection{The Kerr-de Sitter spacetime}
As said above the Kerr-de Sitter spacetime can be regarded as a particular case of the Wahlquist solution just discussed.
The Kerr-de Sitter limit is obtained if we set the restrictions \cite{MARSWAHLQUISTNEWMAN}
$$
Q_0=A^2\;,\quad
\nu_0=1-A^2\Lambda/3\;,\quad
a_1=0\;,\quad
a_2=-2M\;,\quad
\mu_0=-\Lambda\;,\quad
\beta=0\;,
$$
where $A$, $M$ and $\Lambda$ correspond to the angular momentum per unit mass, 
the mass and the cosmological constant of the Kerr-de Sitter spacetime, respectively.
In figure \ref{fig:KerrDesitter}
we show a plot of the quality factor 
$\hat q = \tilde q=\kappa\varkappa$ computed for the Killing vector $\vzeta=\partial/\partial t$.
\begin{figure}
\includegraphics[width=.5\textwidth]{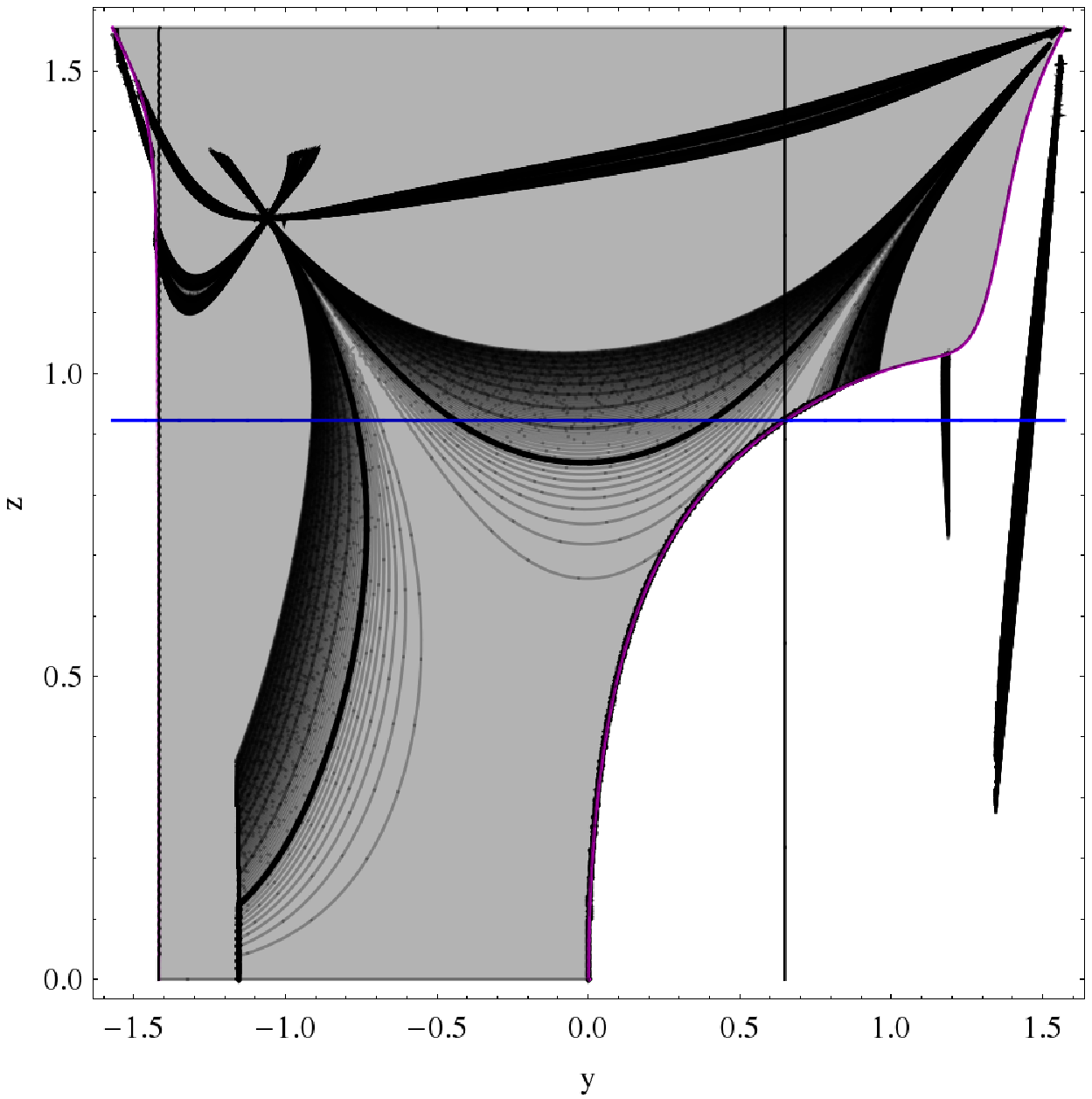}
\includegraphics[width=.5\textwidth]{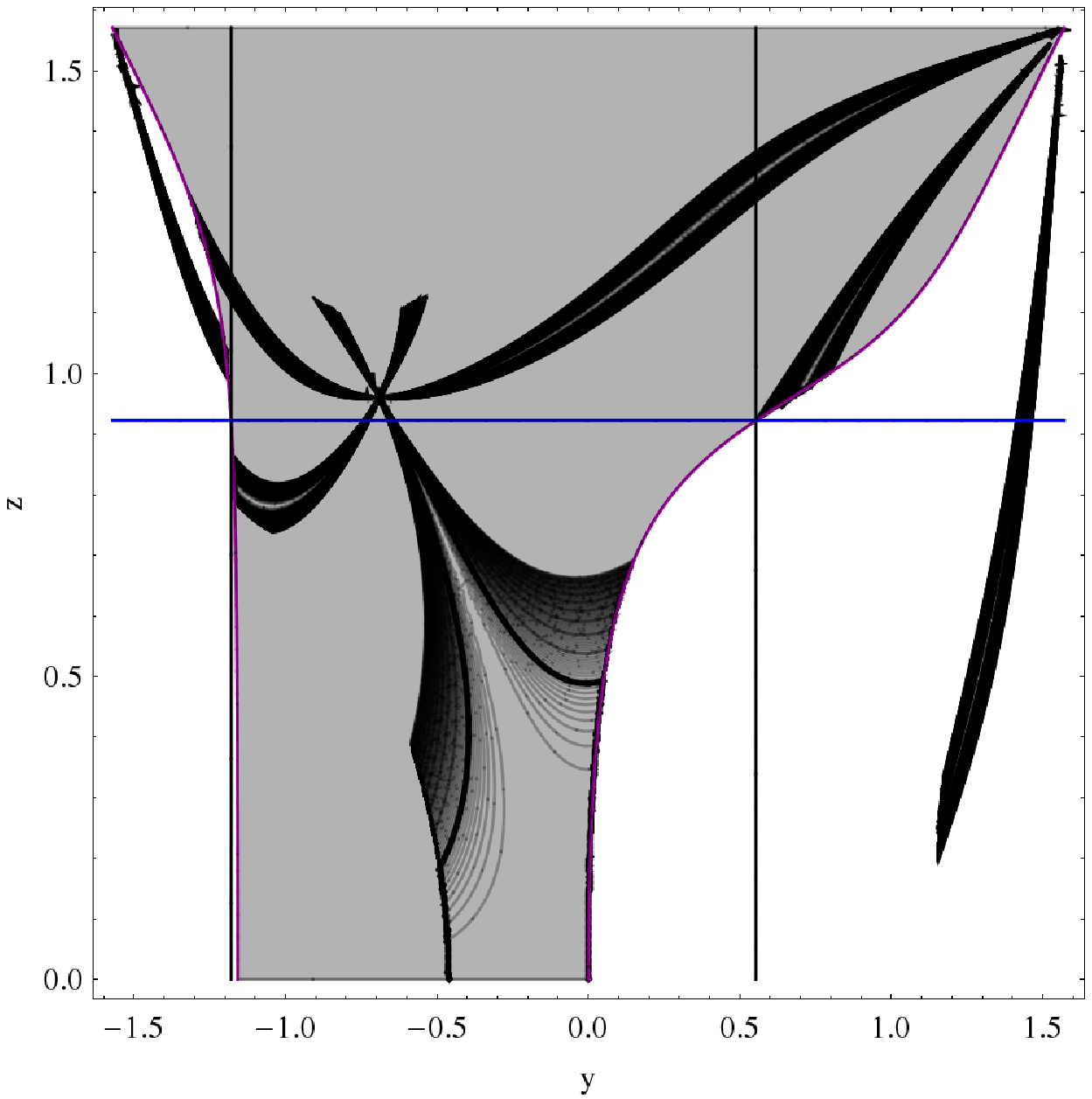} 
\caption{\label{fig:KerrDesitter} In these figures we show a representation of the quality factor 
$\hat q =\tilde q= \kappa\varkappa$ for the Kerr-de Sitter solution with parameters
$M=1.5$, $A=1.32$ and Killing vector $\vzeta=\partial/\partial t$. In the left picture $\Lambda=0.1$ and 
in the right picture $\Lambda=1$. We have represented
the quality factor in the plane $y-z$ by contour curves (contour interval $1/1000$) and only curves with values 
greater than $99/100$ are represented (this is the region we regard as ``close'' to unity). The coordinates have been 
compactified through the relations $y\rightarrow\tan y$,
$z\rightarrow\tan z$. In the graphs, $y\in(-\pi/2,\pi/2)$, $z\in(0,\pi/2)$ (the part with $z\in(-\pi/2,0)$ 
is obtained by a reflection under the axis $z=0$). We have also included a representation of the set of points 
in which the Killing vector $\partial/\partial t$ is timelike (shaded region) and the points where 
the determinant spanned by the Killing vectors $\partial/\partial t$, $\partial/\partial\phi$ is vanishing 
(vertical lines and horizontal line). There is a region where the quality factor is above $999/1000$ which is
bounded by the thick contour line (it contains the singularity $y=0,z=0$ as an accumulation point). 
This region is larger in the left picture as $\Lambda$ is smaller in this case. 
One can check that the size of this region increases when $\Lambda$ approaches zero.}
\end{figure}

\subsection{The Kerr-Newman spacetime}
Consider the Kerr-Newman spacetime in the standard Boyer-Linquidst coordinates
\begin{eqnarray}
&& g_{tt}=\frac{2 MR-A^2\cos^2\theta-Q^2-R^2}{R^2+A^2\cos^2\theta}\;,\quad
g_{t\phi}=-\frac{A(2 M R-Q^2)\sin^2\theta}{R^2+A^2\cos^2\theta}\;,\nonumber\\
&& g_{\theta\theta}= R^2+A^2\cos^2\theta\;,\quad g_{RR}=\frac{R^2+A^2\cos^2\theta}{A^2+Q^2+R^2-2 MR}
\nonumber\\
&& g_{\phi\phi}=\frac{\sin^2\theta\left((A^2+R^2)^2-A^2(A^2+Q^2+R^2-2 M R)\sin^2\theta\right)}{R^2+A^2 \cos^2\theta}\;,
\end{eqnarray}
which is a solution of the Einstein-Maxwell equations.
For the Kerr-Newman spacetime the Ernst 1-form is not closed and hence one cannot define the Ernst potential as used here. 
However, as is well known \cite{EXACTSOLUTIONS}, one can define a complex potential ---in general 
Einstein-Maxwell spacetimes--- for a linear combination of the twist one-form $\omega_c$ with 
a one-form associated to the complex self-dual Maxwell 2-form and the Killing vector. 
Using this generalized potential one can define generalizations of the Mars-Simon tensor 
and find characterizations of the Kerr-Newman space-time analogous to those used in this paper for the 
Kerr case: see \cite{BJM2,WONG} for details. Given that the Kerr-Newman space-time 
is asymptotically flat one can thereby define a quality factor completely analogous to $q$. 
This is, however, outside the scope of this paper.

On the other hand, we can pursue our program and compute the quality factors $\tilde q q_\sigma$ and $\hat q q_\sigma$ for the Kerr-Newman solution. It turns out that 
$q_1=q_2=1$ as one can explicitly check, and therefore, the quality factor which we need to study is 
$\tilde q q_\sigma= \hat q q_\sigma= q_{\sigma}\kappa\varkappa$. In this case one can compute the factors $\kappa$ and $\varkappa$ 
in a closed form getting 
\begin{eqnarray}
&&\kappa=\left\{\begin{array}{c}
             +1\;,\quad  \beta+\lambda=-1+\frac{Q^2}{A^2 \cos^2\theta+R^2}<0\;,\\
              0\;,\quad \beta+\lambda=-1+\frac{Q^2}{A^2 \cos^2\theta+R^2}>0\;.
               \end{array}\right.\\
&&\varkappa=\left\{
\begin{array}{c}
\frac{\left(A^2 M \cos^2\theta+R \left(MR-Q^2\right)\right)^2}{\left(A^2\cos^2\theta+R^2\right) 
\left(A^2 M^2 \cos^2\theta+\left(Q^2-M R\right)^2\right)}\;,\quad  
\mbox{sign}\left(\boldsymbol{g}\left(\frac{\partial}{\partial t},\frac{\partial}{\partial t}\right)\right)<0\;,\\
\frac{A^2 Q^4 \cos^2\theta}{\left(A^2\cos^2\theta+R^2\right) \left(A^2 M^2\cos^2\theta+
\left(Q^2-M R\right)^2\right)}\;,\quad
\mbox{sign}\left(\boldsymbol{g}\left(\frac{\partial}{\partial t},\frac{\partial}{\partial t}\right)\right)>0.
\end{array}\right.
\end{eqnarray}
Note that in the limit $Q\rightarrow 0$, $\varkappa$ goes to $1$ only outside the ergosphere.
This does not contradict the  proposition \ref{prop:varkappa} as it is formulated 
under the assumption ${\boldsymbol g}(\partial/\partial t,\partial/\partial t)<0$. 
In the region  ${\boldsymbol g}(\partial/\partial t,\partial/\partial t)>0$, $\varkappa$ 
tends to $0$ when $Q$ approaches zero. 
A numerical study
of $q_{\sigma}\kappa\varkappa$ for some selected cases is shown 
in figure \ref{fig:KerrNewmann}.

\begin{figure}
\includegraphics[width=.5\textwidth]{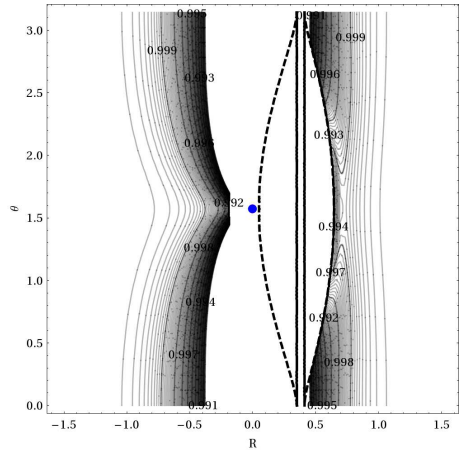}
\caption{\label{fig:KerrNewmann} Contour curves for the quality factor 
$\tilde q q_\sigma= \hat q q_\sigma= q_{\sigma}\kappa\varkappa$ for the Kerr Newman solution with 
$\vzeta=\partial/\partial t$. The coordinate $R$ (horizontal axis) has been compactified with the transformation $R\rightarrow\tan R$
and the coordinate $\theta$ runs along the vertical axis (hence the upper and lower parts of the plots represent the axis 
of the Killing field $\partial/\partial\phi$ and they should be identified). 
The values of the mass, angular momentum per unit mass and charge are, respectively, $M=0.4$, $A=0.35$ and $Q=0.19$ 
and in this case there are two Killing horizons represented 
by the two parallel vertical lines. The dashed curves are the ergosurfaces and the dot is the ring singularity 
located at $R=0$, $\theta=\pi/2$. Only contour curves with values of the quality factor greater than $99/100$ have been represented
(the contour interval is $1/10000$).  The quality factor approaches unity for 
$R\rightarrow\pm\pi/2$ (the asymptotically flat regions).}
\end{figure}

\subsection{The Lense-Thirring approximation}
The Lense-Thirring metric can be regarded as a first order approximation to the (exterior) metric of a rotating compact body. 
The metric tensor components are obtained from
(\ref{eq:asymptotic-flat-end}) by just retaining the leading terms
\begin{equation}
 g_{tt}=-1+\frac{2M}{r}\;,
\quad g_{tx^i}=-\epsilon_{ijk}\frac{4 S^ix^k}{r^3}\;,\quad
g_{x^ix^j}=\delta_{ij}.
\label{eq:lense-thirring}
\end{equation}
Without loss of generality we may adopt the choice $S^1=S^2=0$ (i.e. we choose a coordinate system adapted to the body rotation axis).
For our computations, it is convenient to express the above metric in spherical coordinates $(R,\theta,\phi)$ which are related to the coordinates
$(x^1,x^2,x^3)$ through the usual relation $x^1=R\sin\theta\cos\phi$, $x^2=R\sin\theta\sin\phi$, $x^3=R\cos\theta$.
We present in figure \ref{fig:lense-thirring} a numerical study of the quality factor $q_\sigma\hat{q}$ for a certain choice of 
$M$ and of the non-vanishing component $S^3$.
\begin{figure}
\includegraphics[width=.5\textwidth]{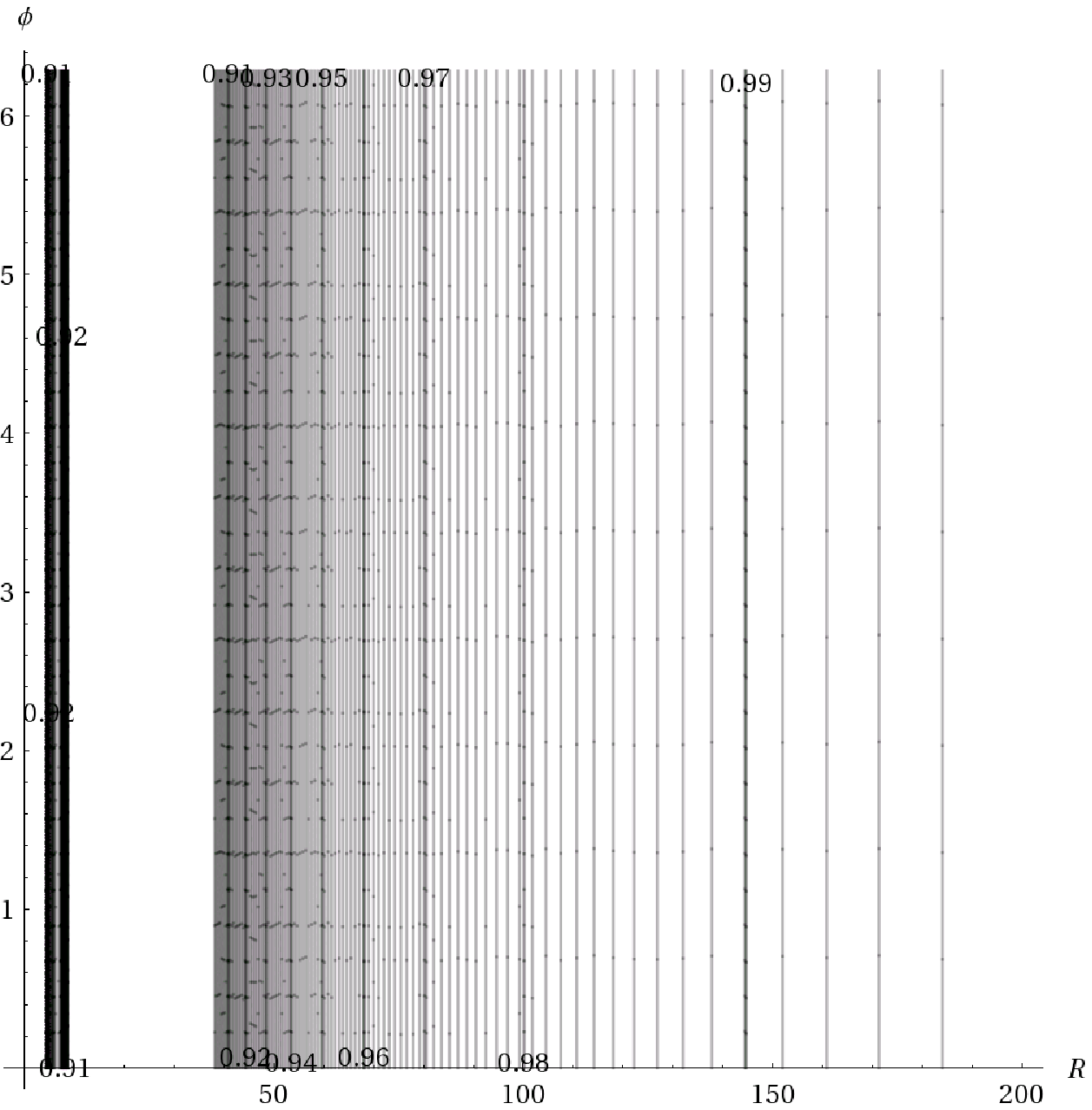}
\includegraphics[width=.5\textwidth]{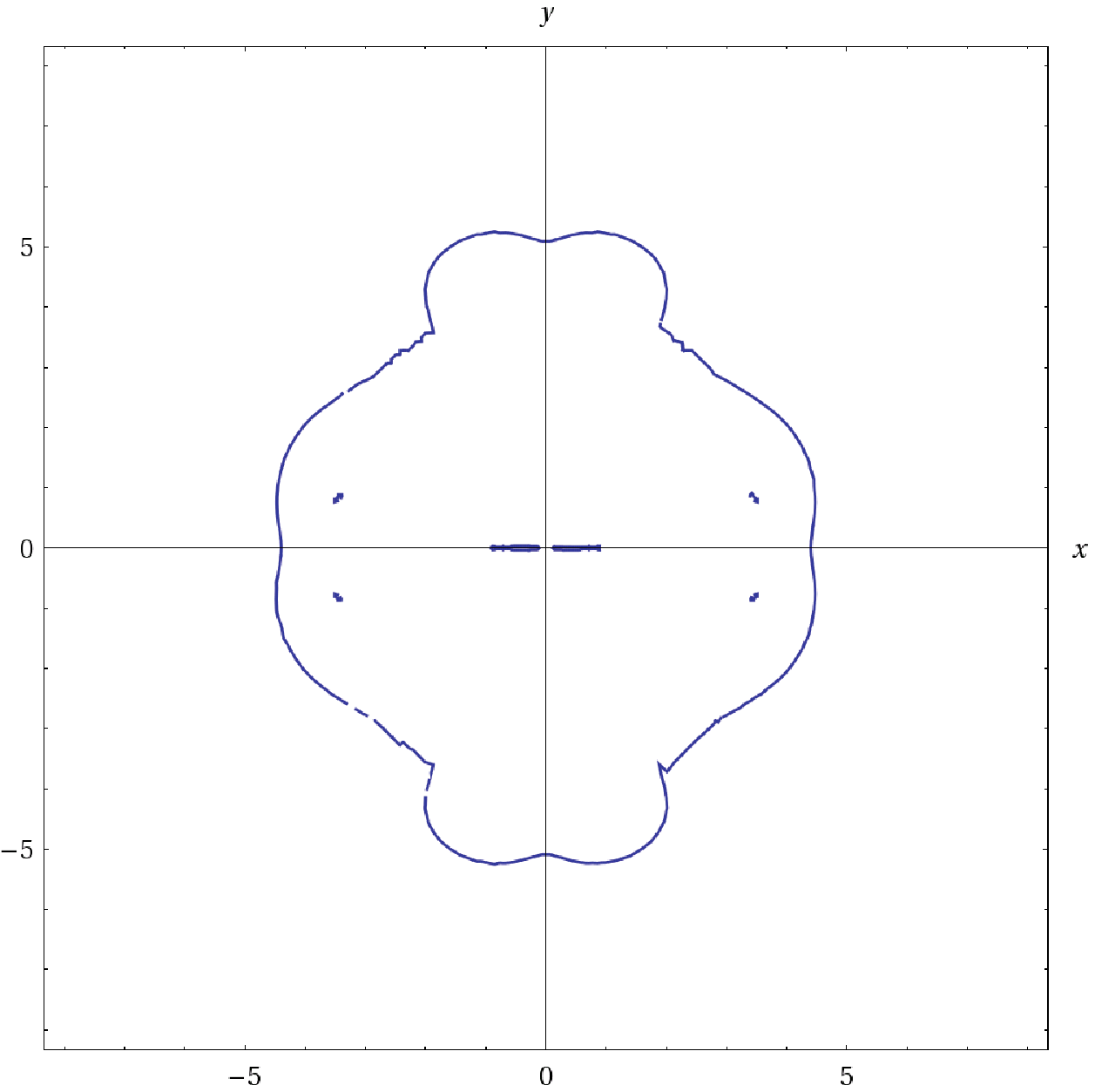}
\caption{\label{fig:lense-thirring} The graph on the left is a plot of the contour curves for the quality factor 
$\hat{q}$ for the case of the approximate Lense-Thirring solution with parameters $M=2$, $S^1=S^2=0$, $S^3=14$ 
and fixed $\theta=\pi/4$ ($t$ is suppressed so only the $R-\phi$ dependency is kept). 
The contour interval is $1/1000$ and only values of the quality factor greater than $0.9$ are represented. 
Note that the curves are all vertical lines and one can check that this behaviour is maintained for different values of $\theta$, 
suggesting that the contour hypersurfaces have axial symmetry, as expected on intuitive grounds. 
We represent one of these contour hypersurfaces
in the graph on the right. In this graph the coordinates $x^2$ and $t$ have been suppressed and the value of $\hat{q}$ is $8/10$. 
Due to the axial symmetry one just has to rotate the curve drawn in the graph around the 
vertical line to obtain the corresponding contour hypersurface.}
\end{figure}

\section{Conclusions}
\label{sec:conclusions}
Taking the invariant characterizations of the Kerr solution written in terms of the Mars-Simon and the space-time Simon tensors we have constructed a number of dimensionless, non-negative scalar quantities (quality factors) which attain the value
+1 in a Ricci flat space-time if and only if it is locally isometric to the Kerr solution. These quality factors depend on the existence 
of a time-like Killing vector in the space-time and as such they can only be used on stationary space-times. 
However, we must stress that in all the examples studied in section \ref{sec:examples} the positivity properties of the quality factors 
were kept even in those points where the Killing vector is not time-like. This strongly suggests that all the quality factors studied
in this paper are indeed non-negative regardless of the causal character of the Killing vector entering in their definition. A rigorous proof
of this fact has eluded us so far.  

Another interesting question regarding the quality factors is whether they fulfill some hyperbolic differential equation
similar to eq. (\ref{eq:covdiv-ms}) for the Mars-Simon tensor. If that was the case then one could analyse the Cauchy problem of 
such an equation and the global existence of its solutions. This could serve to address the non-linear stability of the Kerr solution 
under the assumption of the existence of a Killing vector.  

We have already mentioned in the introduction the uniqueness results dealing with the Kerr solution and the recent progress in this direction.
In \cite{IONESCU-KL-ALEXAKIS} a uniqueness result for the Kerr solution was proven which applies to smooth
asymptotically flat stationary vacuum solutions of the Einstein equations.
Besides the usual assumptions of the space-time being the exterior
region of a regular black hole, a hypothesis about the space-time being ``close'' to Kerr is assumed. This ``Perturbation of Kerr assumption''
states that the components of the Mars-Simon tensor with respect to an orthonormal frame are bounded by a certain quantity when evaluated on a 
partial Cauchy hypersurface of the {\em domain of outer communication}. One could entertain the possibility of expressing this assumption 
in terms of one of the quality factors introduced in this article. This might result in alternative, maybe more powerful, uniqueness results for the Kerr solution.

Finally, there is the open question of how to generalize the quality factors to the generic situation without Killing vector fields.

\begin{acknowledgements}
We thank Dr. Marc Mars for relevant comments on an earlier version of this manuscript.
JMMS is supported by grants
FIS2010-15492 (MICINN), GIU06/37 (UPV/EHU), P09-FQM-4496 (J. Andaluc\'{\i}a---FEDER) and UFI 11/55 (UPV/EHU).
AGP is supported by the Research Centre of Mathematics of the University of
Minho (Portugal) through the ``Funda\c{c}\~ao para a Ci\^encia e a Tecnolog\'{\i}a (FCT) Pluriannual
Funding Program'' and through project CERN/FP/123609/2011.
\end{acknowledgements}


%
%

\end{document}